\documentclass[aps, preprint,nofootinbib,preprintnumbers,eqsecnum,superscriptaddress,sort]{revtex4}
\pdfoutput=1
\usepackage{amssymb,amsmath,amsfonts,amsthm,latexsym,mathrsfs}
\usepackage{graphics, graphicx,color,epsfig,subfigure}
\usepackage[
      colorlinks=true,
      linkcolor=blue,
      urlcolor=blue,
      filecolor=black,
      citecolor=red,
      pdfstartview=FitV,
      pdftitle={},
        pdfauthor={},
        pdfsubject={},
        pdfkeywords={},
        bookmarksopen=true,
      ]{hyperref}

\numberwithin{equation}{section}
\def \be {\begin{equation}}
\def \ee {\end{equation}}
\def \bea {\begin{eqnarray}}
\def \eea {\end{eqnarray}}

\newcommand{\ka}{\kappa}

\newcommand{\bln}{\begin{align}}
\newcommand{\eln}{\end{align}}
\newcommand{\bst}{\begin{split}}
\newcommand{\est}{\end{split}}
\newcommand{\bi}{\begin{itemize}}
\newcommand{\ei}{\end{itemize}}
\newcommand{\ben}{\begin{enumerate}}
\newcommand{\een}{\end{enumerate}}

\def\le{\left}
\def\ri{\right}
\def\ha{{1\over 2}}

\def\al{{\alpha}}

\def\tr{{\rm tr}}
\def\Tr{{\rm Tr}}

\def\th{{\theta}}

\def\ep{{\epsilon}}

\newcommand{\p}{\partial}

\newcommand\ga{{\ensuremath{{\gamma}}}}
\newcommand\Ga{{\ensuremath{{\Gamma}}}}

\newcommand\sig{\sigma}
\newcommand\Sig{\Sigma}
\newcommand\lam{\lambda}
\newcommand\Lam{\Lambda}
\newcommand\om{\omega}
\newcommand\Om{\Omega}

\def\lam{{\lambda}}

\def\eeq{\end{equation}}

\def\Tr{\mathop{\rm Tr}}

\newcommand\sA{{\ensuremath{{\mathcal A}}}}
\newcommand\sB{{\ensuremath{{\mathcal B}}}}
\newcommand\sC{{\ensuremath{{\mathcal C}}}}

\newcommand\sL{{\ensuremath{{\mathcal L}}}}
\newcommand\sM{{\ensuremath{{\mathcal M}}}}

\newcommand\sO{{\ensuremath{{\mathcal O}}}}

\newcommand\sJ{{\mathcal J}}

\newcommand\sR{{\ensuremath{{\mathcal R}}}}

\newcommand\bpsi{{\bar \psi}}

\def\th{{\theta}}

\newcommand\Gab{C}

\begin{document}
\title {Anomalies of the Entanglement Entropy in Chiral Theories}

\author{Nabil Iqbal}
\email{n.iqbal@uva.nl}
\affiliation{Institute for Theoretical Physics, University of Amsterdam, Science Park 904, Postbus 94485, 1090 GL Amsterdam, The Netherlands}

\author{Aron C. Wall}
\email{aroncwall@gmail.com}
\affiliation{
School of Natural Sciences,
 Institute for Advanced Study,
 Princeton, New Jersey 08540, USA}
\vspace{1 cm}

\begin{abstract}\noindent{
We study entanglement entropy in theories with gravitational or mixed $U(1)$ gauge-gravitational anomalies in two, four and six dimensions. In such theories there is an anomaly in the entanglement entropy: it depends on the choice of reference frame in which the theory is regulated.  We discuss subtleties regarding regulators and entanglement entropies in anomalous theories.
We then study the entanglement entropy of free chiral fermions and self-dual bosons and show that in sufficiently symmetric situations this entanglement anomaly comes from an imbalance in the flux of modes flowing through the boundary, controlled by familiar index theorems.

In two and four dimensions we use anomalous Ward identities to find general expressions for the transformation of the entanglement entropy under a diffeomorphism. (In the case of a mixed anomaly there is an alternative presentation of the theory in which the entanglement entropy is not invariant under a $U(1)$ gauge transformation. The free-field manifestation of this phenomenon involves a novel kind of fermion zero mode on a gravitational background with a twist in the normal bundle to the entangling surface.) We also study $d$-dimensional anomalous systems as the boundaries of $d+1$ dimensional gapped Hall phases. Here the full system is non-anomalous, but the boundary anomaly manifests itself
in a change in the entanglement entropy when the boundary metric is sheared relative to the bulk. }
\end{abstract}
\maketitle

\vspace{-10pt}
\tableofcontents
\section{Introduction}

Quantum field theories can have anomalies.  These are subtleties which arise when the regulator of the theory breaks some of the symmetries which were preserved by the classical version of the theory.  The goal of this paper is to describe an anomaly in the entanglement entropy which appears in certain chiral field theories.

Entanglement entropy is a hot topic in high energy physics, condensed matter physics, and black hole thermodynamics.\footnote{For some reviews, see \cite{Calabrese:2005zw, Amico:2007ag, Ryu:2006bv, Solodukhin:2011gn}.}  Formally, we can define the the entanglement entropy $S$ on any region $A$ of a Cauchy slice $\Sigma$ of a spacetime, by evaluating the von Neumann entropy of the density matrix $\rho_A$ of the fields restricted to $A$:
\be
S = -\mathrm{tr}(\rho_A \ln \rho_A)
\ee
However, the entanglement entropy in QFT is UV divergent due to the entanglement of short-distance degrees of freedom across the boundary $\partial A$ (called the ``entangling surface'').  Because of this, $S$ depends not only on the choice of region $A$, but also on the regulator scheme used to cut off the short distance entanglement.  This can lead to an unexpected anomalous transformation of $S$ under a symmetry for which it was na\"{i}vely invariant.

\begin{figure}[h]
\begin{center}
\includegraphics[scale=0.5]{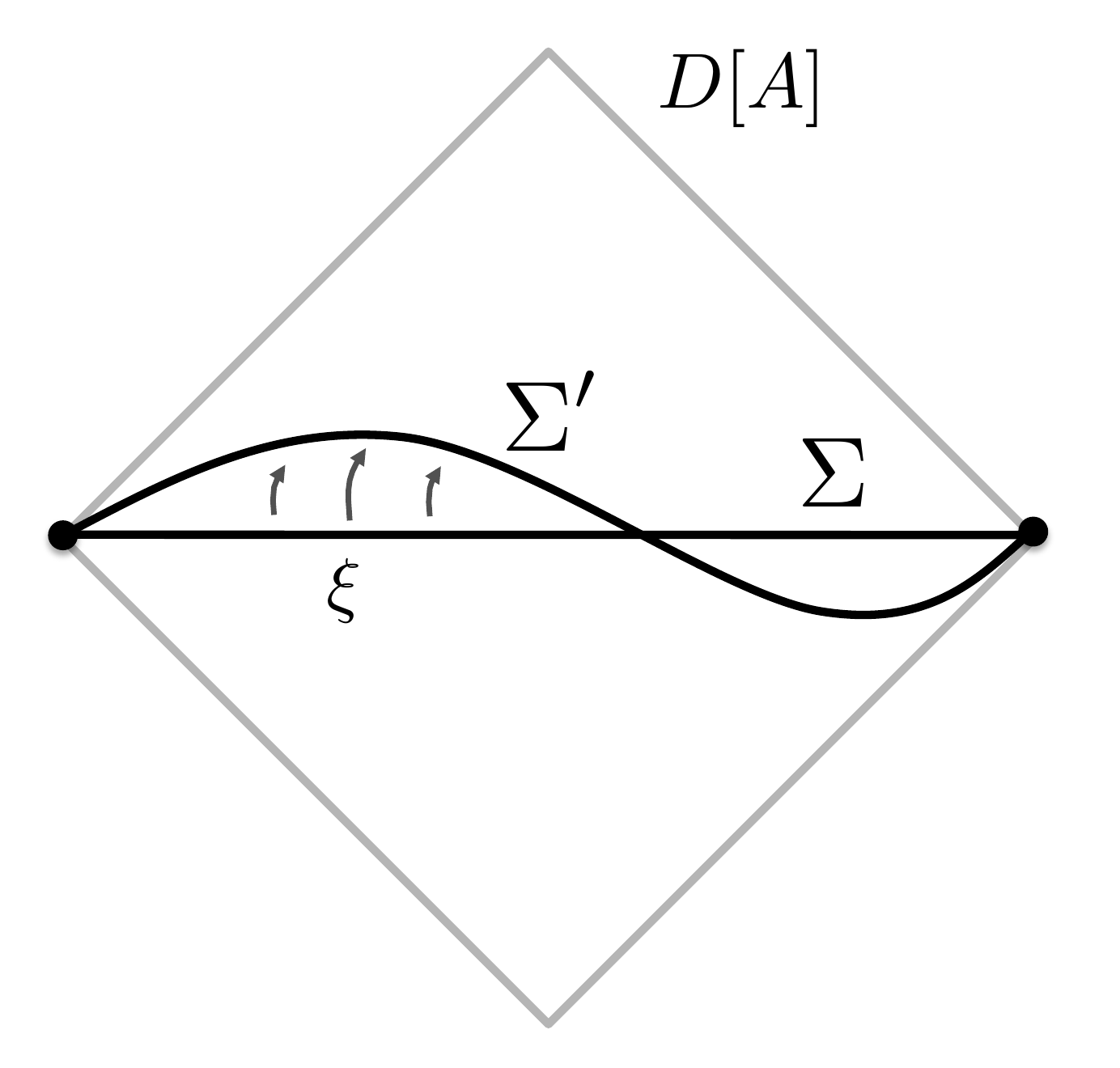}
\end{center}
\vskip -0.5cm
\caption{Two different Cauchy slices $\Sigma$ and $\Sigma'$ of the same domain of dependence $D[A]$, connected by a diffeomorphism $\xi$.}\label{fig:causaldep}
\end{figure}

Normally, $S$ has the property that it depends only on the domain of dependence $D[A]$ of the region $A$.  One would have expected that any partial Cauchy slice $\Sigma$ of $D[A]$ would have the same amount of entropy on it, because the information on two such slices $\Sigma$ and $\Sigma'$, as shown in Figure \ref{fig:causaldep}, are related by a unitary transformation, which preserves the von Neumann entropy.\footnote{This is a formal argument.  From an algebraic perspective, the transformations we consider are actually outer automorphisms of the algebra of observables.  (An ``outer automorphisms'' is a symmetry of the algebra which does not correspond to any well-defined unitary operators in the algebra of $A$, modulo ``inner automorphisms'' which do correspond to unitaries.)  From this perspective, the possible existence of an anomaly is not completely surprising.}  However, in a theory with a diffeomorphism anomaly, this property no longer holds.

This entanglement anomaly only appears in quantum field theories which also have a diffeomorphism anomaly.  Although the diffeomorphism anomaly makes it impossible to define the theory on a general curved spacetime (without adding additional structure), one might have thought that such theories are perfectly well-behaved in flat spacetimes.\footnote{Or more generally, on a curved spacetime for which the diffeomorphism anomaly vanishes.}  However, if one wishes to calculate the entanglement entropy in such theories, one finds that it depends on the reference frame in which one regulates the theory, so that the entanglement entropy is not preserved by a local Lorentz boost.  In 2 dimensions this ambiguity was first pointed out in \cite{Wall:2011kb}, and discussed from a dual holographic perspective in \cite{Castro:2014tta}. In this paper we will describe the 2d anomaly from several complimentary points of view, and also extend the results to 4 and 6 spacetime dimensions.

In $D \ge 4$ dimensions, there can be a mixed anomaly which involves both a $U(1)$ gauge field and the gravitational field.  For simplicity we focus on $D = 4$.  In this case there is a free parameter, which can be adjusted to decide whether the theory should break gauge invariance or diffeomorphism invariance (or both).  This choice determines the invariance properties of the entanglement entropy.  When diffeomorphism symmetry is broken, we find that the entanglement entropy transforms under a local boost, in the presence of a magentic flux through $\partial A$.  On the other hand, when gauge symmetry is broken, the entanglement entropy transforms under a local gauge transformation, in the presence of a gravitational ``twist'' field along $\partial A$.\footnote{Although this gravitational anomaly is not present in the Standard Model, that is only because of a cancellation beween various chiral fermions.  The entanglement entropy of an individual chiral field is thus still ambiguous under a boost or gauge transformation.  See the Discussion section.}

In dimensions of the form $D = 4k + 2$, there exists a purely gravitational anomaly.  In $D = 6$, we will show that there is a boost anomaly in the entanglement entropy which can arise when $\partial A$ has a nonzero Pontryagin number. 
 We expect similar results to hold in higher dimensions.

Some other recent articles on this topic are \cite{Loga,NY}\footnote{We thank T. Azeyanagi, R. Loganayagam, G-S. Ng, T. Nishioka, and A. Yarom for correspondence and for sharing their drafts with us prior to publication.}. \cite{Loga} studies the problem from the dual AdS/CFT point of view, but on general grounds their results for the frame-dependence of the entanglement entropy should apply to any field theory. Our results, derived in a rather different manner, are in precise agreement with theirs when a comparison is possible. \cite{NY} uses methods similar to ours, but treats the presence of coordinate singularities differently, resulting in a factor of two disagreement in various expressions. See also \cite{Guo:2015uqa,Hosseini:2015uba,Belin:2015jpa} for further investigation into and applications of holographic entanglement entropy in theories with anomalies. 

\subsection{Analogy to trace anomaly}
A more familar example of an anomaly in the entanglement entropy comes in CFT's.  Here, scale invariance maps a spatial region $A$ to a rescaled region $A'$, for example one twice as large.  Thus one might have expected that $S(A) = S(A')$.  But in fact this is not the case, because $S$ can depend on the ratio between the length scale of the region and the UV cutoff.  Thus $S$ transforms in an anomalous way under rescaling; the naive scale invariance is not present.  But this is so disturbing if we think of the CFT as an effective field theory description of a microscopic theory which in fact has a shortest distance scale.

This non-scale invariance of the entanglement entropy is closely related to the \emph{trace anomaly}, which is a nonzero trace $T$ of the stress tensor which arises when a CFT is quantized on a curved spacetime.  This anomaly exists in even numbers of dimensions, e.g. in 4 dimensions the trace anomaly (of a theory without a diffeomorphism anomaly) takes the form
\be
T = -aE_4 + cC^2
\ee
where $E_4 = R_{abcd}R^{abcd} - 4 R_{ab} R^{ab} + R^2$ is the Euler density and $C^2 = C_{abcd} C^{abcd} = R_{abcd}R^{abcd} - 2 R_{ab} R^{ab} + (1/3)R^2$ is the Weyl-squared invariant, and $a$ and $c$ are the central charges of the theory.  The interesting thing is that these same coefficients occur in the divergence structure of the entanglement entropy, for example in 4d \cite{Fursaev:2013fta,Solodukhin:2008dh}:
\be 
S(R) = \# \frac{\mbox{Area}}{\epsilon^2} - 8\pi \int_{\partial A} \sqrt{h} d^2x [-a E_2 + c I] \ln(\epsilon) + \mathrm{finite}
\ee
where $\epsilon$ is a UV cutoff, $\#$ is a nonuniversal number, $h$ is the 2 dimensional metric,
\be
E_2 = R[h]
\ee
is the 2d Euler density of the boundary, and
\be
I = R_{ijkl}h^{ik}h^{jl} + R_{ij}h^{ij} + (1/3)R - K_{ij} K^{ij} + (1/2)K^2
\ee
is another conformally invariant density involving both Riemann and the extrinsic curvature.  The coefficient in front of power law divergences such as the area term is nonuniversal, meaning that it depends on the details of the UV regulator.  These power law divergences can be subtracted off in a canonical way which does not require picking a length scale; they are therefore unrelated to the physics of anomalies.  But the coefficient of the log divergence is universal, and what is more it transforms additively under a multiplicative change of scale.  It is therefore not surprising that it is related to the physics of the trace anomaly, as discussed in \cite{Holzhey:1994we,Ryu:2006ef,Solodukhin:2008dh,Hung:2011xb,Schwimmer:2008yh,Casini:2011kv}.

The connection to the trace anomaly can be made precise by calculating $S$ via the replica trick, described below in section \ref{sec:regulate}.  Because the replica trick involves passing to a spacetime manifold with a curvature singularity at the tip, there is a delta function singularity of $T$ at the tip which causes the path integral to be noninvariant under a local rescaling at the boundary $\partial A$.  The dependence of the log divergence on $a$ and $c$ follows directly from this fact.  (When evaluating the entropy of a Killing spacetime, $E_2(\partial A)$ and $X(\partial A)$ are simply the Wald entropies associated with $E_4$ and $C^2$ respectively.)

Since the replica trick involves passing to a curved spacetime, it stands to reason that the diffeomorphism anomaly, which manifests as a nonconservation of $T_{ab}$, should also manifest as an anomaly in the entanglement entropy.  As stated above, we shall see that the anomaly takes the form of a frame-dependence of the entanglement entropy.

Unlike the trace anomaly, the diffeomophism anomaly appears only in theories with chiral fields (e.g. chiral fermions or self-dual p-form fields).  A purely gravitational anomaly can appear only in spacetime dimensions of the form $D = 4k + 2$ \cite{AlvarezGaume:1983ig}.  But there are also mixed anomaly diagrams which appear for $D = 2k \ge 4$, which cannot be regulated in a way which simultaneously preserves diffeomorphism invariance and gauge invariance.  Thus, if we analyse the theory in the gauge-preserving frame in which diffeomorphism invariance is anomalously broken, there can also be a frame-anomaly in the entanglement entropy.

A pure gauge anomaly would not be relevant, since the gauge potential plays no role in the replica trick calculation of $S$.

In short, the following analogy obtains:
\be
\text{trace anomaly : log divergence of }S\text{ :: chiral anomaly : boost non-invariance of }S.
\ee

\subsection{Regulators and replicas}\label{sec:regulate}

In order to even define the entanglement entropy in an anomalous theory, we need to have a UV regulator for $S$ which permits an anomalous theory.  This is harder than it looks, because several common regulators for the entanglement entropy do not permit chiral fields.

Some examples of regulators that do not work: (1) A lattice regulator makes the von Neumann entropy well defined (although there are subtleties for lattice gauge theories \cite{Buividovich:2008gq, Donnelly:2011hn, Casini:2013rba}), but is subject to the fermion doubling problem, resulting in an non-anomalous theory. We discuss this further below. (2) An t' Hooft brick wall \cite{'tHooft:1984re} just outside the entangling surface would require some kind of reflecting boundary conditions to be placed on the brick wall; but in an anomalous theory such boundary conditions are not possible because the number of left and right moving modes can be different (as we shall see explicitly in sections \ref{sec:flow}, \ref{sec:4dweyl}, and \ref{sec:6dindex} below.)

\begin{figure}[h]
\begin{center}
\includegraphics[scale=0.9]{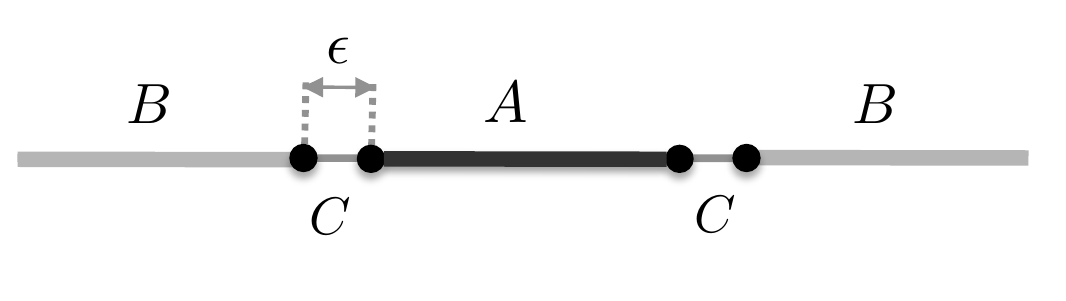}
\end{center}
\vskip -0.5cm
\caption{Example of mutual information regulator. All intervals are understood to be at the same time slice.}\label{fig:mutualinfo}
\end{figure}

{\bf Mutual Information.} One regulator which \emph{does} work for chiral theories involves a limit of the mutual information \cite{Casini:2004bw}.  The first step is to widen the entangling surface $\partial R$ into a region $C$ with small width of order $\epsilon$.  This defines a slightly smaller region $A \subset R$, similarly there is a slightly smaller region inside the complement $B \subset \bar{R}$, as shown in Figure \ref{fig:mutualinfo}.  For example $A$ and $B$ could be the set of points whose distance from $\partial R$ on the time slice $\Sigma$ is greater than $\epsilon$.  Then one can define the mutual information
\be
I(\epsilon) \equiv I_{A,B} = S(\rho_{AB} | \rho_A \otimes \rho_B) = S_A + S_B - S_{AB},
\ee
in terms of the relative entropy $S(\rho|\sigma) \equiv S = \mathrm{tr}(\rho \ln \rho) - \mathrm{tr}(\rho \ln \sigma)$.  Here the first equality is the algebraic definition of the mutual information, which is well-defined and finite in any reasonable QFT, so long the minimum gap between $A$ and $B$ is finite.  The second inequality holds for finite systems whose entropy is well-defined.  In a QFT $S_A$, $S_B$, and $S_{AB}$ are separately divergent (and difficult to define in a chiral theory), but the divergences are local on the boundary, and therefore cancel between the three terms.  Thus \emph{formally} we may say that the second inequality holds as well in QFT.

For a spin system on a discrete lattice, if we set $\epsilon = 0$ so that $A$ and $B$ are complimentary regions, then in a pure state $S_A = S_B$ and $S_{AB} = 0$; hence
\be
I_{A,B} = 2S_{A} \label{IisS}
\ee
In a continuum QFT, $I(\epsilon)$ diverges in the limit that $\epsilon \to 0$, and so we cannot set $\epsilon = 0$.  Nevertheless, motivated by \ref{IisS}, we may define $I(\epsilon)/2$ at small but finite $\epsilon$ as a regulated version of the entropy $S(A)$, in a pure state.  This is a strictly formal relation, as always with regulators; nevertheless one can see that it is reasonable by observing that the contributions from long range entangled entities (e.g. EPR pairs) are the same for both $I(\epsilon)/2$ and $S(A)$.

Note that unlike the lattice or brick wall, the mutual information regulator is purely passive, in the sense that it does not modify the physics, only the definition of $S$.  However, it does depend on the choice of slice $\Sigma$.  So it is not manifest that $I(\epsilon)$ is independent of the reference frame used to define $A,B,C$.  And in fact we shall see that in theory with a chiral diffeomorphism anomaly, this non-boost-invariance actually arises.

{\bf Replica Trick.}  A final way to define the entanglement entropy is by means of the replica trick \cite{Holzhey:1994we, Calabrese:2004eu, Calabrese:2005zw, Cardy:2007mb}.  In this trick, we first Wick rotate to a Euclidean manifold which generates the state, and then pass to the $n$-fold cover of this Euclidean manifold, so that there is a conical singularity with angle $2 \pi n$ going around the entangling surface $\partial A$.  The partition function $Z_n$ of this replicated manifold is related to the Renyi entropy $\tr(\rho^n)$.  We can if we like regulate this conical singularity by smoothing it out over a distance $a$.  We also need a UV regulator $\epsilon$ on the field theory, in order to make it finite.  Finally we analytically continue
\be
\frac{1}{1-n}\ln\left(\frac{Z_n}{Z_1^n}\right)
\ee
to $n = 1$ in order to obtain the regulated von Neumann entropy $S_n$.  Typically when we do this, there are divergences as $\epsilon \to 0$ but not as $a \to 0$ \cite{Nelson:1994na}.

In the case of a theory with a gravitational anomaly, this procedure becomes trickier.  We will not spend too much time worrying about exactly how to impose the UV regulator $\epsilon$, since the form of the anomaly itself should be independent of the regulator.  But it is conceptually important that the regulator, whatever it is, must break coordinate invariance.  One manifestation of this is that the stress-tensor $T_{ab}$\footnote{Here we refer to the ``consistent'' form of $T_{ab}$ obtained by varying $\ln Z$ with respect to the metric} depends on the Christoffel symbol $\Gamma$.  This means that the theory implicitly requires a coordinate system (with its associated flat auxilliary Cartesian metric) in order for it to be well-defined.

Now in order for the replica trick to make sense, all physically relevant structures must be faithfully replicated $n$ times (except near the tip which may be smoothed out to regulate the answer).  Hence, the auxilliary flat coordinate system must itself be copied $n$ times.  This leads to a coordinate singularity at $\partial R$.  But this coordinate singularity remains even after the curvature singularity is smoothed out.\footnote{In 2 dimensions, this impossibility of smoothing the coorindate singularity can easily be seen by choosing a unit timelike vector associated with one particular coordinate, say $\hat{x}$, and then observing that $\hat{x}$ is twisted around $\partial R$ the wrong number of times to have a smooth interior.}

It is therefore necessary to define the theory even in the presence of a coordinate singularity.  This requires one to specify boundary conditions at the singularity, i.e. one must specify the state which pops out of it.  In a 2 dimensional CFT with a scale-invariant coordinate singularity at the origin, it is most natural to assume that in radial quantization, the state coming out of the singularity should be the dilaton vacuum.

More generally, we may argue that if the theory is tensored with its $P$-inverse theory, the resulting $\text{QFT}_L \times \text{QFT}_R$ has no gravitational anomaly.  It is therefore well-defined even in the presence of a coordinate singularity.  Let the state coming out of the singularity be
\be
\Psi = \Psi_L \otimes \Psi_R
\ee
where the state factorizes into $\text{QFT}_L$ and $\text{QFT}_R$ modes, because the two sectors do not interact.  But this defines $\Psi_L$ up to a phase.  (Assuming the singularity is parity symmetric, this phase must be real and hence a sign.)  In any case, if we take on faith that the theory is well-defined in the presence of the coordinate singularity, we can define the variation of $S$ with respect to a boost and get a definite answer; we will perform this calculation in sections \ref{sec:2dpathint}, \ref{sec:4dpathint}.  So long as one is careful to take into account the divergences of $T_{ab}$ near the coordinate singularity, one obtains the same answer by the replica trick path integral as by other methods.

\subsection{Regularizing with an extra dimension}

There is another, rather different way to regularize an anomalous theory. If a lattice regulator for a given quantum field theory exists then we are guaranteed to have a well-defined (if non-universal) notion of entanglement entropy for a given spatial region.\footnote{We note that the simplest implementation of this idea requires modification in the case of lattice gauge theory\cite{Buividovich:2008gq,Donnelly:2011hn,Casini:2013rba} where the physical Hilbert space does not factor across lattice sites, but the idea itself still makes sense.  See also \cite{Ohmori:2014eia,Donnelly:2014fua,Donnelly:2015hxa,Donnelly:2014gva,Huang:2014pfa} for a discussion of related issues from the point of view of the continuum field theory.}  As mentioned above, however, anomalous theories generally can {\it not} be regulated by a lattice in the UV while preserving the symmetries of the problem. This is the essential content of the fermion-doubling theorems of Nielsen and Ninomiya \cite{Nielsen:1980rz,Nielsen:1981xu}. In such anomalous theories it seems that there is then a certain difficulty in precisely localizing degrees of freedom in space. In much of this paper, we will ignore this subtlety and proceed with path integral computations using the replica trick, but one might rightfully question whether the entanglement entropy that we so compute necessarily has a Hilbert space interpretation.

However, there is a related system which {\it does} have a lattice regulator. Anomalous $d$-dimensional quantum field theories {\it can} be understood as living on the boundary of a $(d+1)$-dimensional gapped field theory with some topological structure. The simplest example of this is perhaps the theory of a single right-moving Weyl fermion in $(1+1)$ dimensions, which exhibits a two-dimensional gravitational anomaly and can be understood as the edge mode of a traditional integer quantum Hall droplet (see e.g. \cite{wen1992theory,fradkin2013field} for introductory reviews). The combined bulk + boundary system is invariant under $d+1$ dimensional diffeomorphisms -- in a sense the anomalies of the two theories ``cancel'' -- and so can be realized with a lattice regulator. Thus it should be possible to unambiguously define the entanglement entropy in this system, though it may be difficult to separate entanglement of the topological bulk from that of the gapless edge modes. We stress that there is no notion of duality being used here (or indeed anywhere in this paper); the bulk and boundary {\it simultaneously} exist.

However one might wonder whether the entanglement entropy in such a system exhibits any signature of the anomaly at all. As it turns out, we can frame the non-invariance of the entanglement entropy discussed above in terms of such a system, where we consider an entangling region that extends into the bulk, and then study the response of the system under a deformation of the bulk metric that ``shears'' the boundary relative to the bulk, pulling it infinitesimally in the direction of a specified $d$-dimensional vector field that can be thought of as a ``boundary diffeomorphism''. Just as above, the response of the entanglement to such an operation is given by a local integral over the entangling surface. The form of this integral is again completely fixed by the anomaly, though it turns out to be related to its {\it covariant} rather than its consistent form, as we discuss in detail later in Sections \ref{sec:2dpathintbdy} and \ref{sec:4dpathintbdy}.

This can be viewed as yet another regulator on an anomalous theory, a particularly uneconomical one that requires the presence of an entire extra dimension.

\subsection{Plan of paper}

The introduction being nearly over, tradition dictates that we warn our readers of the things which are to come.

In each section $I < D < VII$ of the body of the paper, we will discuss the physics of a $D$-dimensional system of physics related to the anomaly.  When $D$ is even, this means that we will describe the chiral diffeomorphism or mixed gauge-gravitational anomaly in $D$ dimensions, and calculate its effects on a boost or gauge transformation of the entanglement entropy.  Since there is no anomaly in odd dimensions, for odd $D$ we will instead discuss a $D = d+1$ dimensional Hall system, which has a $d$ dimensional anomalous theory living on its boundary and provides another perspective on the anomaly in the entanglement entropy.  In section VII we will discuss.

\section{Gravitational anomaly in two dimensions}\label{sec:2d}

In this section we will consider a CFT$_2$ which has a gravitational anomaly, i.e. $c_L \ne c_R$.\footnote{Such a CFT can still be modular-invariant if $c_L - c_R$ = 0 mod 24 and other conditions are met.  The simplest examples of such theories involve 24 chiral bosons compactified on a Niemeier lattice (one of the 24 different even unimodular lattices possible in 24 dimesnsions), although other constructions such as orbifolding give additional theories \cite{Schellekens:1992db}.  We will not require modular invariance for what follows.}

We will show in three different ways that even if we do not couple the theory to gravity, there is still a residual effect of this anomaly, namely that the entanglement entropy of a region is not invariant under boosting the cutoff.  This is not too surprising given that many techniques for calculating the entanglement entropy involve passing to a curved spacetime \cite{Holzhey:1994we,Calabrese:2009qy,Calabrese:2004eu}. 

In theories with $c_L = c_R$, the gravitational anomaly cancels, but the fact that the left and right-moving sectors are separately anomalous implies that entropy is shifted between them under a boost.  Thus there is no invariant notion of the entropy of just the left-movers.

We first review some elementary aspects of the form of the gravitational anomaly \cite{AlvarezGaume:1983ig}. This material is well-known \cite{Bardeen:1984pm,AlvarezGaume:1984dr,Ginsparg:1985qn}. We have benefitted from the reviews provided by the recent treatments in e.g. \cite{Jensen:2012kj,Jensen:2013kka,Jensen:2013rga}. In Euclidean signature we define the stress tensor of a two-dimensional field theory with partition function $Z[g] \equiv e^{-W[g]}$ as the response to an infinitesimal variation of the metric:
\be
T^{\mu\nu}(x) \equiv \frac{2}{\sqrt{g}}\frac{\delta W[g]}{\delta g_{\mu\nu}(x)}  \label{Tdef}
\ee
If the theory suffers from a gravitational anomaly then this stress tensor is not conserved:
\be\label{stressncons}
\nabla^{\mu}T_{\mu\nu} =-i c_g \epsilon^{\rho \sig}\p_{\rho}\p_{\mu}\Gamma^{\mu}_{\nu \sig}~
\ee
with $c_g$ an anomaly coefficient. The factor of $i$ on the left-hand side is due to the Euclidean signature; it is only the imaginary part of a Euclidean partition function that can be anomalous, but upon analytic continuation to Lorentzian signature typically all sensible observables are real, as we will explicitly see.

If we are studying a two-dimensional conformal field theory then the anomaly coefficient can be related to a difference in central charges as follows:
\be
c_g \equiv \frac{c_L - c_R}{96\pi} \ .
\ee
Now we note that the right-hand side of the anomaly equation \eqref{stressncons} does not appear covariant; there is an explicit appearance of the Christoffel symbol. Relatedly, the object $T^{\mu\nu}$ defined as a functional derivative of a generating function in \eqref{Tdef} does not actually transform as a tensor. This is called the ``consistent'' form of the anomaly, as it satisfies the Wess-Zumino consistency condition \cite{Bardeen:1984pm}.

We can define a covariant stress tensor by adding a local functional of the sources, usually called the Bardeen-Zumino counterterm \cite{Bardeen:1984pm}:
\be
T^{\mu\nu}_{cov} \equiv T^{\mu\nu} + T^{\mu\nu}_{BZ} \label{covdef2d}
\ee
where explicitly the counterterm is
\be
T^{\mu\nu}_{BZ} \equiv -\ha \nabla_{\lam}\le(X^{\lam\mu\nu} + X^{\lam\nu\mu} - X^{\mu\nu\lam}\ri) \qquad X^{\mu\lam}_{\phantom{\mu\lam}\nu} \equiv - i c_g \le(\ep^{\mu\rho}\Ga^{\lam}_{\phantom{\lam}\nu\rho} + \ep^{\lam\rho} \Ga^{\mu}_{\phantom{\mu}\nu\rho}\ri) \label{BZdef}
\ee
The resulting $T^{\mu\nu}_{cov}$ transforms as a tensor, and its divergence is covariant:
\be
\nabla_{\rho}T^{\rho\mu}_{cov} =  -i c_g \ep^{\mu\nu}\nabla_{\nu}R, \label{covanom}
\ee
However it is not the functional derivative of a two-dimensional action. We will revisit this point later.

Finally, we note that we can always trade a diffeomorphism anomaly for a Lorentz anomaly, in which case $T^{\mu\nu}$ remains conserved but develops an antisymmetric piece. This is physically equivalent and in this work we will focus on the presentation of the anomaly in which the stress tensor is not conserved.

\subsection{Physical entanglement flow argument}\label{sec:flow}

There is an easy geometrical way to see the anomaly in $D = 2$ \cite{Wall:2011kb}.  The entanglement entropy of an interval of length $L$ is given by
\be\label{EEL}
S = \frac{c_L + c_R}{12} \ln\left(\frac{L^2}{\epsilon_1 \epsilon_2}\right) + s(\rho),
\ee
where $s(\rho)$ is a finite, state dependent contribution and there is a cutoff (e.g. the mutual information regulator of section \ref{sec:regulate} which cuts of the UV divergence at a proper distance $\epsilon_1, \epsilon_2$ along the left and right hand sides of the interval respectively.\footnote{If the theory has a parity symmetry $c_L = c_R$.  Since there are two endpoints of $L$, so we recover the standard $S = (c/3)\ln(L) + \mathrm{const.}$ vacuum entanglement formula \cite{Holzhey:1994we}.}

\begin{figure}[h]
\begin{center}
\includegraphics[scale=0.5]{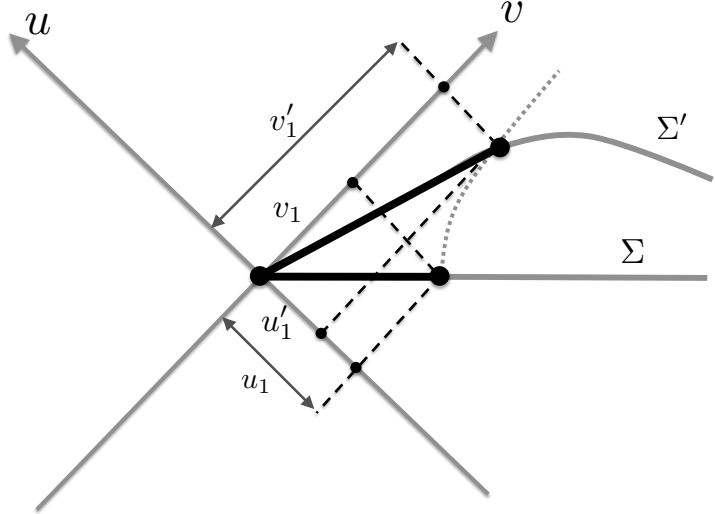}
\end{center}
\vskip -0.5cm
\caption{Zoomed in view near one of the endpoints of the interval, demonstrating transformation of the cutoff under a local boost. $(u_1, v_1)$ and $(u_1', v_1')$ refer to the lengths of the cutoff along the $(u,v)$ directions before and after the boost respectively.}\label{fig:geometric}
\end{figure}

Now $L$ has a domain of influence $D[L]$.  But we can also choose a different Cauchy slice $\Sigma$ of $D[L]$ with the same endpoints, and impose the cutoff in the reference frame of this slice, as shown in Figure \ref{fig:geometric}.  Na\"ively, the two slices have the same information, and so it makes no difference to the von Neumann entropy $S$.  However, if there is a nontrivial boost angle at either of the two endpoints, information can leak in and out past the cutoff surface, so there is the possibility for an anomaly.

The key realization is that $c_L$ is due to degrees of freedom which propagate leftwards at the speed of light, while $c_R$ is due to degrees of freedom which propagate rightward at the speed of light.  This tells us that the entropy transforms under a boost by an amount proportional to $c_L - c_R$.  To find the exact constant of proportionality, we can split \eqref{EEL} into left and right moving pieces:
\be
S = \frac{c_L}{12} \ln\left(\frac{\Delta v^2}{v_1 v_2}\right) + \frac{c_R}{12} \ln\left(\frac{\Delta u^2}{u_1 u_2}\right) + s(\rho),
\ee
where $v_{1,2}$ and $u_{1,2}$ measure the proper length of the cutoff along the $v$ and $u$ directions at endpoints $1$ and $2$.
If the slice $\Sigma$ is boosted at the endpoints by a Lorentzian angle $\chi_{1,2}$ relative to $L$, then we have
\begin{align}
v_i &=  e^\chi_i \epsilon, \\
u_i &= e^{-\chi_i}\epsilon.
\end{align}
Here we have defined a positive boost to be one for which $\delta t / \delta x$ is positive (note that this definition is not preserved under reflecting the $x$ coordinate), and thus
\be
\delta S = \frac{c_R - c_L}{12} \left[\delta \chi_1 + \delta \chi_2\right] \label{Sans}
\ee
Hence, if we bulge the slice $\Sigma$ towards the future in a reflection symmetric way, we obtain the same entropy.  But in general, different Cauchy slices will have different amounts of entanglement entropy on them, after subtracting off the divergences using the cutoff.  The logarithmic entanglement divergence acts as a Hilbert hotel, allowing one to increase or decrease the amount of entropy by means of the boost.  Thus, as in the case of other anomalies, a classically true conservation law is violated by the quantum field theory.

Note that although the entanglement entropy depends on the boost angle at which the slice $\Sigma$ hits the boundary, the entanglement $S$ is still invariant under a \emph{global} Lorentz boost of flat spacetime, assuming the boost also acts on the interval (and associated slice $\Sigma$).  That is because we are regulating the entanglement entropy in the reference frame associated with the slice $\Sigma$.  We could instead have selected a unit timelike vector ``aether field'' $u^a$, and regulated the entanglement in the reference frame associated with this field.  This makes the entropy of an interval independent of $\Sigma$, but breaks global Lorentz symmetry through the choice of aether field.  It seems likely that this is conceptually related to trading the diffeomorphism anomaly for a Lorentz violation anomaly, but we will not explore this relationship more here.

\subsection{Connection with Casimir energy}
There is a slicker way to derive the entanglement anomaly for a CFT, by exploiting the Casimir momentum on a cylinder. To do this, we start by conformally transforming the $n$-fold place to a cylinder using the map $z = e^{iw}$, where $z$ is the complex Cartesian coordinate on the plane and $z = 0$ is the entangling surface.  This gives us a cylinder whose radius is $R = 2\pi n$.  The ground state of the cylinder corresponds to the dilaton vacuum in radial coordinates.
As a result of the conformal transformation, there is an anomalous change in the stress-tensor, which can be derived from the Schwarzian via standard arguments.  One finds that the Casimir energy is
\be
E = -2\pi\frac{c_L + c_R}{24R}.
\ee
Since left-moving fields move left, and right-moving fields move right, it follows that when $c_L \ne c_R$ there is also a Casmir momentum \cite{Kraus:2005zm,Jensen:2012kj}:
\be
p = 2\pi\frac{c_L - c_R}{24R},
\ee
so that the vacuum state on the cylinder is not translation invariant, but instead picks up a phase upon being rotated.

Since the stress-tensor transforms anomalously under this conformal transformation, it is not immediately obvious that a replica trick calculation of $S$ should give the same answers on the cylinder and the plane.  However, for purposes of calculating $\delta S$ it actually is acceptable to use the cylinder frame.  We will first derive the answer and then explain why it works.

If we cut the plane at some radius $|z| = r_*$ and rotate the disk inside by an angle $\theta$ (making $\Gamma \ne 0$ there), on the cylinder this corresponds to cutting at a fixed moment of time and rotating the past by the angle $\theta$.  This generator of this transformation is the Casmir momentum.  Thus the change of the cylinder partition under an infintesimal phase $\delta \theta$ is
\be
\delta \ln Z = ip\, \delta \theta. 
\ee
We then calculate the entanglement entropy $S = (1 - R \partial_R) \ln Z$ and obtain the change of the entropy under a Euclidean rotation:
\begin{align}
\delta S &= \left.(1 - R \partial_R) \delta \ln Z  \nonumber\right|_{R = 2\pi} \\ &= i \frac{c_L - c_R}{12} \delta \theta.
\end{align}
This is imaginary in Euclidean signature, but it corresponds to a real shift in the entropy in Minkowski signature if we act with a real Lorentz boost $\delta \chi = -i \delta \theta$:
\begin{equation}
\delta S = -\frac{c_L - c_R}{12} \delta \chi
\end{equation}

This agrees with the result calculated in the previous section.  This is actually somewhat remarkable, since the partition function $Z$ transforms anomalously under the transformation of the plane to the cylinder.  If instead of calculating both terms in $\delta S$, we had calculated only $\delta \ln Z$, we would \emph{not} have gotten the same answer for both the cylinder and the plane ($Z$ is trivial for the plane, aside from the cosmological constant divergence).

The reason this works is that $S$ transforms in a nicer way than $Z$ does under conformal transformations, since it depends on the UV regulator only near the entangling surface, whereas $Z$ depends on the UV regulator everywhere.  Thus, if we act with a conformal transformation which is trivial in a neighborhood of $z = 0$, but which maps to the cylinder for $|z| > x$ (for some $x$), the entropy is unaffected, and we are free to compute the entanglement entropy at a radial time $r_* > x$ using the calculation above.


\subsection{Path integral derivation} \label{sec:2dpathint}

The previous argument applied to conformal field theories. In this section we present a more general derivation that applies to any 2d field theory (conformal or not) with a gravitational anomaly. We will use the replica approach to the computation of entanglement entropy. We first calculate the Renyi entropy,
\be
S_n = \frac{1}{1-n} \ln \frac{\Tr \rho^n}{(\Tr \rho)^n} \label{renyidef}
\ee
as a Euclidean partition function on an $n$-sheeted Riemann surface with metric $g_{(n)}$ \cite{Calabrese:2009qy,Calabrese:2004eu}. The variation of the partition function on a 2-dimensional manifold $\sM_2$ under an infinitesimal diffeomorphism $\xi$ in any theory is given by
\be
\delta_{\xi} \ln Z[g_{(n)}] = -\ha \int_{\sM_2} d^2x\;\sqrt{g}\langle T^{\mu\nu} \rangle \delta_{\xi}g_{\mu\nu} = - \int_{\sM_2} d^2x \sqrt{g} \;\langle T^{\mu\nu} \rangle \nabla_{\mu}\xi_{\nu}\ .
\ee
In writing this expression we have assumed that the only external source that transforms under diffeomorphisms is the metric. Now when $\sM_2$ has a boundary, this expression can be integrated by parts to obtain
\be
\delta_{\xi} \ln Z[g_{(n)}] = + \int_{\sM_2} d^2x\sqrt{g}\;\nabla_{\mu}\langle T^{\mu\nu} \rangle \xi_{\nu}\ - \int_{\p\sM_2} d\sig\;\sqrt{\ga} T^{\mu\nu}n_{\mu}\xi_{\nu} \label{totdiffvar}
\ee
where the first term measures the intrinsic non-conservation of the stress tensor and the second measures the flow of energy off the edge of the manifold, where $n_{\mu}$ is an outwards-pointing normal to this edge.

In Euclidean space we take the interval to extend from $z_1$ to $z_2$. Let us focus on the neighborhood of $z_1$. We introduce polar coordinates $(r,\th)$ near $z_1$ such that the angular coordinate $\th$ parametrizes rotations around $z_1$ . On the $n$-sheeted Riemann surface $\th$ has periodicity $2\pi n$, and thus $z_1$ is the site of a conical surplus. To discuss physics near that point we may resolve the curvature singularity by introducing a regulated metric $g_{(n),a}$, which near $z_1$ is
\be
ds^2_{(n),a} = f\le(\frac{r}{a}\ri)^2 dr^2 + r^2 d\th^2, \label{regulated}
\ee
with $a$ a small length scale and the function $f(x)$ is chosen to satisfy $f(x \to 0) = n$ and $f(x \to \infty) = 1$. For example, we may take $f(x) = 1 + e^{-x}(n-1)$.

Now consider a general diffeomorphism $\xi$ that corresponds to an infinitesimal and position-dependent rotation on $\theta$ as
\be
\xi = \xi^{\th}(r) \p_{\th} + \xi^{r}(r) \p_r \label{diff1}
\ee
 A short computation evaluating \eqref{stressncons} on the explicit regulated metric \eqref{regulated} shows that the first term in \eqref{totdiffvar} is
\be
\int_{\sM_2} d^2x\;\nabla_{\mu}\langle T^{\mu\nu} \rangle \xi_{\nu} = 4\pi in c_g  \int dr \;\le(\p_r I_a(r)\ri) \xi^\th(r) \qquad I_a(r) \equiv \frac{1}{f^2}\le(\frac{2 r}{a}\frac{f'}{f} - 1\ri) \label{regexp}
\ee
This expression is controlled by $\p_r I_a(r)$. As we take the regulator $a \to 0$, $\p_r I_a(r)$ vanishes for all $r>0$, but its integral from $r$ to $\infty$ is always finite and equal to $n^{-2}-1$. As we take $a \to 0$, the kernel of the integral then becomes a delta function localized at $r = 0$. We find then the following regulator-independent expression for the variation:
\be
\lim_{a \to 0}\int_{\sM_2} d^2x\;\nabla_{\mu}\langle T^{\mu\nu} \rangle \xi_{\nu} = 2\pi i c_g \le(\frac{1}{n} - n\ri)\xi^{\th}(r)\bigg|_{r \to 0}
\ee

We turn now to the boundary term in \eqref{totdiffvar}. At first, it may not appear that the manifold in question has a boundary, as the diffeomorphism dies away at infinity. However, this theory is not generally covariant, and thus it may {\it think} that the origin of polar coordinates -- i.e. the ``circle'' $r = 0$ -- is also a boundary.

To compute the boundary term we need the actual value of the stress tensor near the origin, not just its divergence. Recall now that there is a modified {\it covariant} stress tensor defined in \eqref{covdef2d} that transforms as a tensor. This covariant stress tensor, being covariant, should {\it not} contribute a boundary term from the origin of polar coordinates. For the purposes of obtaining the boundary term we then need only find the contribution from the Bardeen-Zumino correction term, which we can explicitly compute from \eqref{BZdef}:
\be
T^{\mu\nu}n_{\mu} \xi_{\nu}\big|_{r \to 0} \to -T^{\mu\nu}_{BZ} n_{\mu} \xi_{\nu}\big|_{r \to 0} = \frac{3 i c_g \xi^{\th}(r)}{r f\le(\frac{r}{a}\ri)^2}\bigg|_{r \to 0}
\ee
Assembling the pieces we find for the diffeomorphism variation of the $n$-sheeted partition function
\be
\delta_{\xi} \ln Z[g_{(n)}] = -2\pi i c_g\le(\frac{2}{n} + n\ri) \xi^{\th}(r)\bigg|_{r \to 0} \ .
\ee
To find the change in the Renyi entropy we need to also compute the denominator in \eqref{renyidef}, which is simply $n$ times the answer for $n=1$, leaving us with:
\be
\delta_{\xi} S_n = -4\pi i c_g\le(\frac{1}{n} + 1\ri)  \xi^{\th}(r)\bigg|_{r \to 0}
\ee
The dependence on the Renyi index $n$ is familiar from the form of the Renyi entropies for a single interval in 2d CFT. Indeed the transformation of the Renyi entropy under a rigid boost was previously derived from 2d CFT arguments in \cite{Castro:2014tta}. Here we have performed a more general derivation for any diffeomorphism, and we see that the result holds for any two-dimensional field theory (conformal or not) with a gravitational anomaly.

It is clear that we will obtain a contribution from each endpoint of the entangling region $A$, and we may rewrite this result in a more covariant way as
\be
\delta_{\xi} S_n = -2\pi i c_g\le(\frac{1}{n} + 1\ri)\sum_{i \in \p A} \ep^{\mu\nu} \nabla_{\mu} \xi_{\nu}(x_i)
\ee
Finally the entanglement entropy $S$ is related to the Renyi entropy as $S = \lim_{n \to 1} S_n$, leading to
\be
\boxed{\delta_{\xi} S = 8\pi c_g \sum_{i \in \p A}\delta \chi(x_i)} \label{finalans}
\ee
where we have analytically continued to Lorentzian signature in defining the local boost $2\delta\chi(x_i) \equiv i\ep^{\mu\nu}\nabla_{\mu}\xi_{\nu}$. In a CFT this result agrees with the result found earlier on geometric grounds in \eqref{Sans}.

The boundary term representing a flux of stress-energy into the origin was required for this agreement: otherwise the path-integral result here is off by a factor of two. This boundary term may seem unsettling, as it essentially corresponds to a delta function of angular momentum non-conservation that {\it always} remains arbitrarily sharp even though we have smoothened out the curvature singularity over a finite radius $a$. Said differently, while we have regulated the curvature singularity, the coordinate system \eqref{regulated} still has a {\it coordinate} singularity at the origin which we are unable to regulate. In a theory with a diffeomorphism anomaly a coordinate singularity may contribute to physical observables.

In general one might expect a need to specify information at every singularity, indicating a lack of uniqueness from the point of view of the low-energy theory: for example, one might worry that there are other delta functions present whose coefficient we cannot fix from low-energy considerations. In general in quantum field theory it is difficult to rule out the presence of such contact terms: however, examining the structure of the final answer \eqref{finalans} we see that we would require a delta function in $T_{\mu\nu}$ that is antisymmetric. This is not allowed, as we are studying the presentation of the anomaly in which $T_{\mu\nu}$ is symmetric but not conserved. Thus it seems that the anomaly contributes in a universal way, though some care is required with the regulation\footnote{We have also repeated the computation in a Cartesian coordinate system (more precisely, we replicated Cartesian coordinates $n$ times and then smoothed out the metric near the tip): the apportioning of the answer between bulk and delta-function terms is different, but the final answer is the same.}.

\section{Gravitational anomaly on the boundary of a 3d Hall phase} \label{sec:2dpathintbdy}
We now study the anomalous two-dimensional theory as the boundary of a 3d ``Hall'' phase, by which we mean a bulk gapped three-dimensional system that cancels the anomaly of the boundary theory. This is precisely the situation for an ordinary incompressible quantum Hall droplet in the laboratory, where the bulk is made up of some number of Landau levels completely filled with electrons, and the edge mode in question is a single chiral Weyl fermion. This particular system is only an example, and we will not describe its microscopic physics any further, but will simply describe the low-energy effective action describing a general class of such systems\footnote{See also \cite{2012PhRvB..85r4503S} for a discussion on the distinction between covariant and consistent anomalies in relation to condensed matter physics.}. As the full system is diffeomorphism invariant, it admits a lattice regulator and there is no obstruction to defining a microscopic entanglement entropy for spatial subregions in such a system, though it may be hard to separate contributions of the gapped bulk from that of the anomalous boundary.

Denote by $G_{MN}$ the metric of the 3d bulk, and denote by $\Gab^{M}_{PQ}$ its metric-compatible Christoffel connection. The 2d metric $g_{\mu\nu}$ should be understood as describing the boundary of the three-manifold with metric $G_{MN}$. Then the generating functional of the full system is (in Euclidean signature)
\be
W_{tot}[G] = W[g] + i c_g S_{CS}[\Gab]  \label{fullac}
\ee
Here $W[g]$ is the generating functional of the boundary two-dimensional theory as studied above. $S_{CS}[C]$ is the three-dimensional gravitational Chern-Simons term,
\be
S_{CS}[\Gab] \equiv \int_{\sM_3} d^3x\;\overline{\ep}^{MNP} \le(\Gab^{A}_{QM} \p_N \Gab^{Q}_{AP} + \frac{2}{3}\Gab^{A}_{QM} \Gab^{Q}_{BN} \Gab^{B}_{QP}\ri)
\ee
(here $\overline{\ep}$ with an overbar indicates the Levi-Civita {\it symbol}, and an $\ep$ with no overbar denotes the Levi-Civita tensor). This Chern-Simons term is {\it not} the action itself of fundamental degrees of freedom: rather these gapped bulk degrees of freedom have been integrated out, leaving a response functional that captures the response to changes in the fixed external metric $G$.

Consider now a three-dimensional diffeomorphism $\chi$ that acts on $G$ as
\be
\delta_{\chi} G_{MN} = D_{M}\chi_N + D_N \chi_M,
\ee
with $D$ the three-dimensional covariant derivative with respect to connection $\Gab$. Note that $\chi$ need not vanish at the boundary, and will thus induce a transformation of the boundary metric $g$. The Chern-Simons term is invariant under diffeomorphisms up to a boundary term, which precisely cancels the intrinsically two-dimensional anomalous variation of $W[g]$. Then the combined partition function $W_{tot}[G]$ is by construction {\it invariant} under $\chi$:
\be
\delta_{\chi} W_{tot}[G] = 0 \label{3ddiff}
\ee
Recall from \eqref{Tdef} that the usual two-dimensional stress tensor is $T^{\mu\nu} \equiv \frac{2}{\sqrt{g}} \frac{\delta W[g]}{\delta g_{\mu\nu}}$.
This stress tensor satisfies the anomaly equation \eqref{stressncons}. Now consider the change in $W_{tot}$ under a small variation of the metric $\delta G$, which includes a potential variation of the boundary metric $\delta g$. The full variation can be split into several parts:
\be
\delta_{G} W_{tot}[G] = \ha\int_{\p\sM_3} d^2 x \sqrt{g}\le(T^{\mu\nu} + T_{BZ}^{\mu\nu}\ri) \delta g_{\mu\nu} - i c_g\int_{\sM_3} d^3x \sqrt{G} C^{MN} \delta G_{MN} \ . \label{totvar}
\ee
The last term arises from the bulk variation of the gravitational Chern-Simons term \cite{Deser:1981wh}. $C^{MN}$ is called the Cotton tensor, and is
\be
C^{MN} \equiv \ha\le(\ep^{QPM} D_{P}  R_{Q}^N + \ep^{QPM} D_{P} R_{Q}^M\ri) \label{cottondef}
\ee
The first term comes from the usual variation of the boundary 2d field theory. The second term arises from a boundary term coming from the variation of the Chern-Simons term, and is in fact equal to the Bardeen-Zumino correction term defined in \eqref{BZdef}. Recall from \eqref{covdef2d} that the sum of the original stress tensor and the Bardeen-Zumino correction term is the covariant stress tensor $T^{\mu\nu}_{cov}$ whose non-conservation is given by a covariant expression \eqref{covanom}. Note that we are giving the bulk Hall phase a physical interpretation, but it is also often used simply as a technical device to construct the covariant stress tensor.

We stress that \eqref{3ddiff} does not mean that $T_{cov}^{\mu\nu}$ is conserved. Rather it means that the non-conservation of $T^{\mu\nu}_{cov}$ can be interpreted as a flow of energy from the bulk onto the boundary.

We would now like to study how the entanglement entropy in this system responds to two-dimensional diffeomorphisms. It may not be clear what we mean by this: after all, under three-dimensional diffeomorphisms \eqref{3ddiff} shows that the full system is invariant. Nevertheless there is a natural sense in which the system responds to {\it two}-dimensional diffeomorphisms.

\begin{figure}[h]
\begin{center}
\includegraphics[scale=0.5]{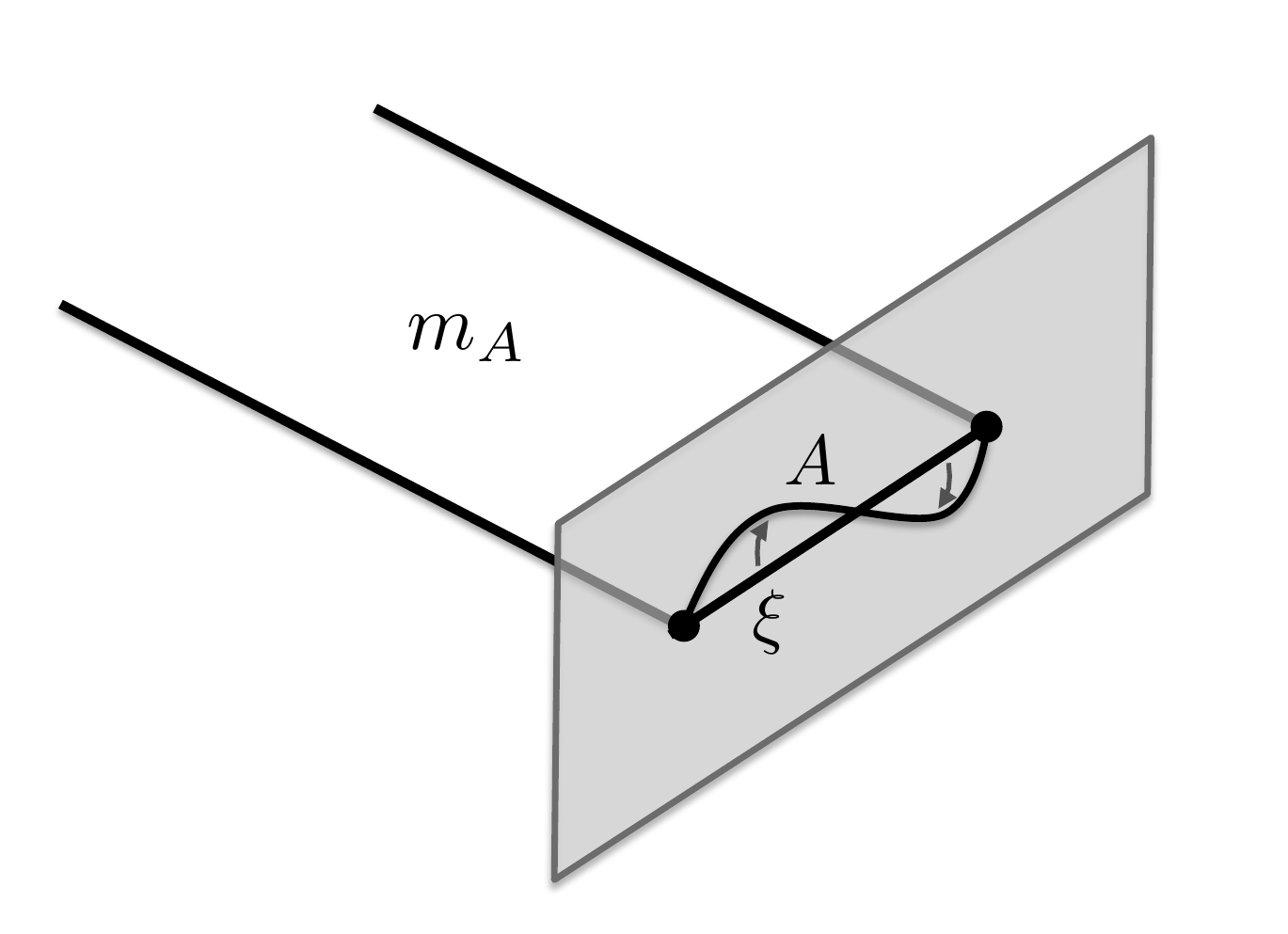}
\end{center}
\vskip -0.5cm
\caption{Boundary region $A$ and associated bulk region $m_A$. Under the described shear operation of the full metric, the interval on the boundary is distorted as shown.}\label{fig:bulkbdy}
\end{figure}

Take the 2d field theory to live on flat 2d space, which is then the boundary of a half-space $\sM_3$. The details of the geometry of $\sM_3$ should not matter, so for simplicity we take the metric of this flat space to be
\be
ds^2 \equiv G_{MN} dX^M dX^N = g_{\mu\nu}(x^{\mu})dx^{\mu} dx^{\nu}  + dz^2 \label{flatmet}
\ee
with the boundary at $z = 0$ and the deep bulk to be at $z \to \infty$. $g_{\mu\nu}(x)$ is the metric on which the 2d field theory is defined. Now consider a spatial region $A$ -- for simplicity, take it to be an interval -- in the 2d field theory. Extend the endpoints of $A$ straight into the bulk to define a two-dimensional spatial region called $m_A$, as shown in Figure \ref{fig:bulkbdy}. We will study the entanglement entropy of $\sA$.

We now need to specify the action of an infinitesimal 2d diffeomorphism $\xi(x^{\mu})$ on this system. Given such a 2d diffeomorphism with compact support, consider the following 3d metric:
\be
ds^2 \equiv G_{MN}^{(\xi)} dX^M dX^N = \le[g_{\mu\nu} + f(z)\le(\nabla_{\mu} \xi_{\nu} + \nabla_{\nu} \xi_{\mu}\ri)\ri] dx^{\mu} dx^{\nu} + dz^2 \label{shearmet}
\ee
where $f(z)$ is a function that smoothly interpolates from $1$ at $z = 0$ to $0$ at $z = \infty$. It is important to note that \eqref{shearmet} is {\it not} a 3d diffeomorphism of \eqref{flatmet}; rather, one might understand it as physically shearing the boundary at $z = 0$ by a vector field $\xi^{\mu}$ relative to a fixed coordinate system at $z \to \infty$. Note that it {\it is} true that the {\it boundary} metric changes from \eqref{flatmet} to \eqref{shearmet} as though under a standard 2d diffeomorphism. The partition functions and entanglement entropy on $G^{(\xi)}$ will differ from that on $G$, and we will now compute this variation. We will denote by $\overline{\delta}_\xi$ the infinitesimal variation that takes us from \eqref{flatmet} to \eqref{shearmet}, with the overbar reminding us that this variation is {\it not} a diffeomorphism.

We will now compute the entanglement entropy and determine how it changes under the above variation. As above, we will use the replica method approach to the computation of entanglement entropy in two-dimensional field theory. In this section we directly compute the entanglement entropy by evaluating
\be
S  = \le(1 - n \frac{\p}{\p n}\ri) \log Z[G_{(n)}]\bigg|_{n \to 1}.
\ee
as a Euclidean partition function on an $n$-sheeted three-dimensional manifold with metric $G_{(n)}$.  We will take the bulk 3d metric to be uniquely specified by the boundary metric $g_{(n)}$ through the relation \eqref{flatmet}, i.e.
\be
G^{(n)}_{MN} dX^M dX^N \equiv g^{(n)}_{\mu\nu} dx^{\mu} dx^{\nu} + dz^2 \label{bulkcone}
\ee
This corresponds to extending the edges of the boundary interval $A$ straight into the bulk along the $z$ direction. We will compute this only in the $n \to 1$ limit: this amounts to computing a partition function on a metric with a conical singularity and extracting the linear dependence on the opening angle.

Now we compute $\overline{\delta}_{\xi} Z$ by varying off of \eqref{bulkcone} in the manner specified in \eqref{shearmet}. The variation  of the partition function can be parametrized in terms of \eqref{totvar} as
\be
\overline{\delta}_{\xi} W_{tot}[G_{(n)}] =  \ha \int_{\p \sM_3} d^2x \sqrt{g_{(n)}}\;T^{\mu\nu}_{cov} \overline{\delta}_{\xi} g_{\mu\nu} - \int_{\sM_3} d^3x \sqrt{G_{(n)}} C^{MN} \overline{\delta}_{\xi} G_{MN} \label{varxi}
\ee
All geometric quantities are computed on the metric \eqref{bulkcone}.

We first compute the bulk contribution from the Cotton tensor. As \eqref{bulkcone} is a direct product of a 2d conical metric with a line, we see that all bulk geometric quantities will only be non-vanishing if they have legs only in the field theory directions. From the definition of the Cotton tensor in \eqref{cottondef} we see that this implies that $C^{MN}$ must have one index in the $z$ direction, but the variation $\overline{\delta}_{\xi} G_{MN}$ is only in the field theory directions. Thus the bulk contribution vanishes.

We turn now to the first term. By construction, the variation of the boundary metric is that of a 2d diffeomorphism,
\be
\overline{\delta}_{\xi} g_{\mu\nu} = \nabla_{\mu} \xi_{\nu} + \nabla_{\nu} \xi_{\mu} \ .
\ee
We may integrate by parts to obtain
\be
\overline{\delta}_{\xi} W_{tot}[G_{(n)}] = -\int d^2x \sqrt{g_{(n)}}\;\le(\nabla_{\mu} T^{\mu\nu}_{cov}\ri) \xi_{\nu} \label{covint}
\ee
We pause to note the physical interpretation of this formula: the system responds to the manipulation above as though the diffeomorphism is being generated by the {\it covariant} stress tensor, {\it not} the consistent stress tensor as in \eqref{totdiffvar}. There is no contradiction here as we are not studying a 2d partition function. Now using \eqref{covanom} for the divergence of $T^{\mu\nu}_{cov}$ and the well known fact that for a cone with opening angle $2\pi(n-1)$ about the point $x = 0$ the Ricci scalar satisfies $R = 4\pi(n-1) \delta^{(2)}(x)$ (see e.g. \cite{Fursaev:1995ef}), we find the variation in the entanglement entropy to be
\be
\overline{\delta}_{\xi} S = 4\pi i  c_g \sum_{x_i \in \p A}\ep^{\nu\sig}\nabla_{\nu} \xi_{\sig}(x_i) \ .
\ee
Using \eqref{fullac} and analytically continuing to Lorentzian signature using $\ep^{\nu\sig}\nabla_{\nu} \xi_{\sig}(x_i) = -2i\delta\chi(x_i)$, we see that for a local boost we have \eqref{Sans}:
\be
\boxed{\overline{\delta}_{\xi} S = 8 \pi c_g \sum_{x_i \in \p A}\delta \chi(x_i)} \ .
\ee
This formula appears identical to the purely two-dimensional formula \eqref{finalans}, but it is computing something different. It is measuring how the entanglement entropy of a {\it three}-dimensional region -- containing both a gapped bulk and a gapless boundary -- changes if the boundary of the region is physically sheared relative to the deep bulk. Its calculation is also different; it arose from the {\it covariant} form of the anomaly and thus did not involve any extra contributions from the coordinate singularity. Interestingly, the final result is the same. In principle, this formula (unlike \eqref{finalans}) could be verified by an explicit Hilbert space computation involving the microscopic fermionic wavefunctions in a Hall phase.

One may ask what role this extra bulk played in this analysis. In essence its role was really to permit a natural definition of a reference coordinate system, that which lives at the ``other boundary'' at $z \to \infty$. In this computation we are asking how the full system responds when the coordinate system defining the field theory is changed relative to this reference coordinate system.

\section{Mixed gauge-gravitational anomaly in four dimensions}

In this section we extend the above results to four dimensions. In particular, we show that a mixed gauge-gravitational anomaly also results in an entanglement anomaly very similar to that discussed above.

We first review some aspects of the mixed anomaly.
We consider a theory with a $U(1)$ current $j^{\mu}$ and a stress tensor $T^{\mu\nu}$. These can be coupled to background fields $a_{\mu}$ and $g_{\mu\nu}$ in the usual manner, and we have the following expressions:
\be
j^{\mu} \equiv \frac{1}{\sqrt{g}}\frac{\delta W[a,g]}{\delta a_{\mu}} \qquad T^{\mu\nu} \equiv \frac{2}{\sqrt{g}}\frac{\delta W[a,g]}{\delta g_{\mu\nu}}
\ee
The usual Ward identity for the consistent currents with a mixed gauge-gravitational anomaly (with anomaly coefficient $c_m$) in four dimensions in the presence of background fields is
\begin{align}
\nabla_{\mu} j^{\mu} & =  -i\frac{c_m}{4} \ep^{\kappa\sig\al\beta} R^{\nu}_{\phantom{\nu}\lam\ka\sig} R^{\lam}_{\phantom{\lam}\al\beta} \nonumber
\\ \nabla_{\nu} T^{\mu\nu} & = f^{\mu}_{\phantom{\mu}\nu} j^{\nu} + i\frac{c_m}{4}a^{\mu}\le(\ep^{\kappa\sig\al\beta} R^{\nu}_{\phantom{\nu}\lam\ka\sig} R^{\lam}_{\phantom{\lam}\al\beta}\ri) \label{Wardmixedorig}
\end{align}
with $f = da$. The non-conservation of $j^{\mu}$ is the usual statement of the anomaly. The non-conservation of $T^{\mu\nu}$ may require some explanation. It has two parts: the first is familiar from non-anomalous systems as the analog of ``Newton's second law'', simply saying that external gauge fields pulling on charges can dump momentum into the system. The second arises due to the non-conservation of $j^{\mu}$. We will refer to this presentation as the ``diff-preserving frame'', as under a diffeomorphism (which acts both on the background gauge field and metric), we have $\delta_{\xi}W[a,g] = 0$.

Now this can be modified by the addition of a local functional of the sources to the generating functional $W$:
\be
\overline{W}[a,g] \equiv W[a,g] + i c_m \int d^4x \sqrt{g} a_{\mu} K^{\mu} \qquad K^{\al} \equiv \ep^{\al\mu\nu\rho} \le(\Ga^{\delta}_{\mu\sigma} \p_{\nu} \Ga^{\sigma}_{\rho\delta} + \frac{2}{3}\Ga^{\sigma}_{\mu\delta}\Ga^{\delta}_{\nu\ep}\Ga^{\ep}_{\rho\sigma}\ri) \label{counterterm}
\ee
$K^{\al}$ is a four-dimensional analog of the gravitational Chern-Simons term, and its divergence is the gravitational Pontryagain density:
\be
\nabla_{\al} K^{\al} = \frac{1}{4}\ep^{\mu\nu\rho\sig}R^{\al}_{\phantom{\al}\beta\mu\nu}R^{\beta}_{\phantom{\beta}\al\rho\sig}
\ee
Importantly, this counterterm is not diffeomorphism invariant, and so we will no longer have $\delta_{\xi} \overline{W} = 0$. The addition of this counterterm changes the definition of the currents and modifies the Ward identities above to read:
\begin{align}
\nabla_{\mu} \overline{j}^{\mu} & =  0 \nonumber \\
\nabla_{\nu} \overline{T}^{\mu\nu} & = f^{\mu}_{\phantom{\mu}\nu} \overline{j}^{\nu} +  i g^{\mu\nu}\frac{c_m}{2\sqrt{g}}\p_{\lam}\le(\sqrt{g}\ep^{\ka\sig\al\beta}F_{\ka\sig} \p_{\al}\Ga^{\lam}_{\nu\beta}\ri) \label{Wardnice}
\end{align}
We see that we have lost diffeomorphism invariance, but the new current is now conserved, and so we will call this the ``gauge-preserving frame''.

If we wish to study states where we couple a non-trivial background gauge field to the current $j^{\mu}$, then it seems that we should work in the gauge-preserving frame, as defined by \eqref{Wardnice}. We will show below that in the gauge-preserving frame the entanglement entropy exhibits an anomaly under boosts, provided a suitable background gauge field is turned on. On the other hand, if we wish to study the fermions on a non-trivial gravitational background, then it seems that we should work in the diff-preserving frame, as defined by \eqref{Wardmixedorig}. In this case, as one might expect, we will show that the the entanglement entropy transforms anomalously under $U(1)$ gauge transformations. In both cases we will present free fermion computations and general path-integral arguments, just as above.

Finally, it is possible to define a covariant stress tensor and current by adding Bardeen-Zumino improvement terms:
\be
j^{\mu}_{cov} = \overline{j}^{\mu} + \overline{j}^{\mu}_{BZ} \qquad T^{\mu\nu}_{cov} = \overline{T}^{\mu\nu} + \overline{T}^{\mu\nu}_{BZ}
\ee
which take the form
\begin{align}
\overline{T}^{\mu\nu}_{BZ} & \equiv -\ha \nabla_{\lam}\le(X^{\lam\mu\nu} + X^{\lam\nu\mu} - X^{\mu\nu\lam}\ri) \qquad X^{\mu\lam}_{\phantom{\mu\lam}\nu} \equiv \frac{-i c_m}{2}\le(\ep^{\mu\rho\ka\sig} \Ga^{\lam}_{\nu\rho} + \ep^{\lam\rho\ka\sig} \Ga^{\mu}_{\nu\rho}\ri)F_{\ka\sig} \nonumber \\
\overline{j}^{\mu}_{BZ} & \equiv -i c_m K^{\mu} \label{4dBZ}
\end{align}
As above these will play a role in our analysis.

We first study the effect of a magnetic field on the fermion in the gauge-preserving frame.

\subsection{Weyl fermions and chiral zero modes with magnetic fiux} \label{sec:4dweyl}

In this section we review the presence of chiral zero modes when a 4d Weyl fermion is placed in a background magnetic field. This is the essential physics behind the chiral magnetic effect \cite{Fukushima:2008xe}. The relevance of this setup to entanglement entropy has been discussed in a different context in \cite{Swingle:2010bt}.

Consider a single left-handed Weyl fermion in four dimensions with charge $q$. This theory has a gauge-gravitational anomaly with anomaly coefficient $c_m = \frac{q}{192\pi^2}$.\footnote{It also has a $U(1)^3$ anomaly in the charge current, which does not contribute to the entanglement anomaly.  It is most convienent to assume this anomaly is cancelled by other fields, so that we can consistently consider nontrivial bundles for the EM field.  For example, if there are eight left-handed fermions with charge $q = +1$ and one with charge $q = -2$, then the $U(1)^3$ anomaly will cancel, but the mixed anomaly will still be present.}

\begin{figure}[h]
\begin{center}
\includegraphics[scale=0.6]{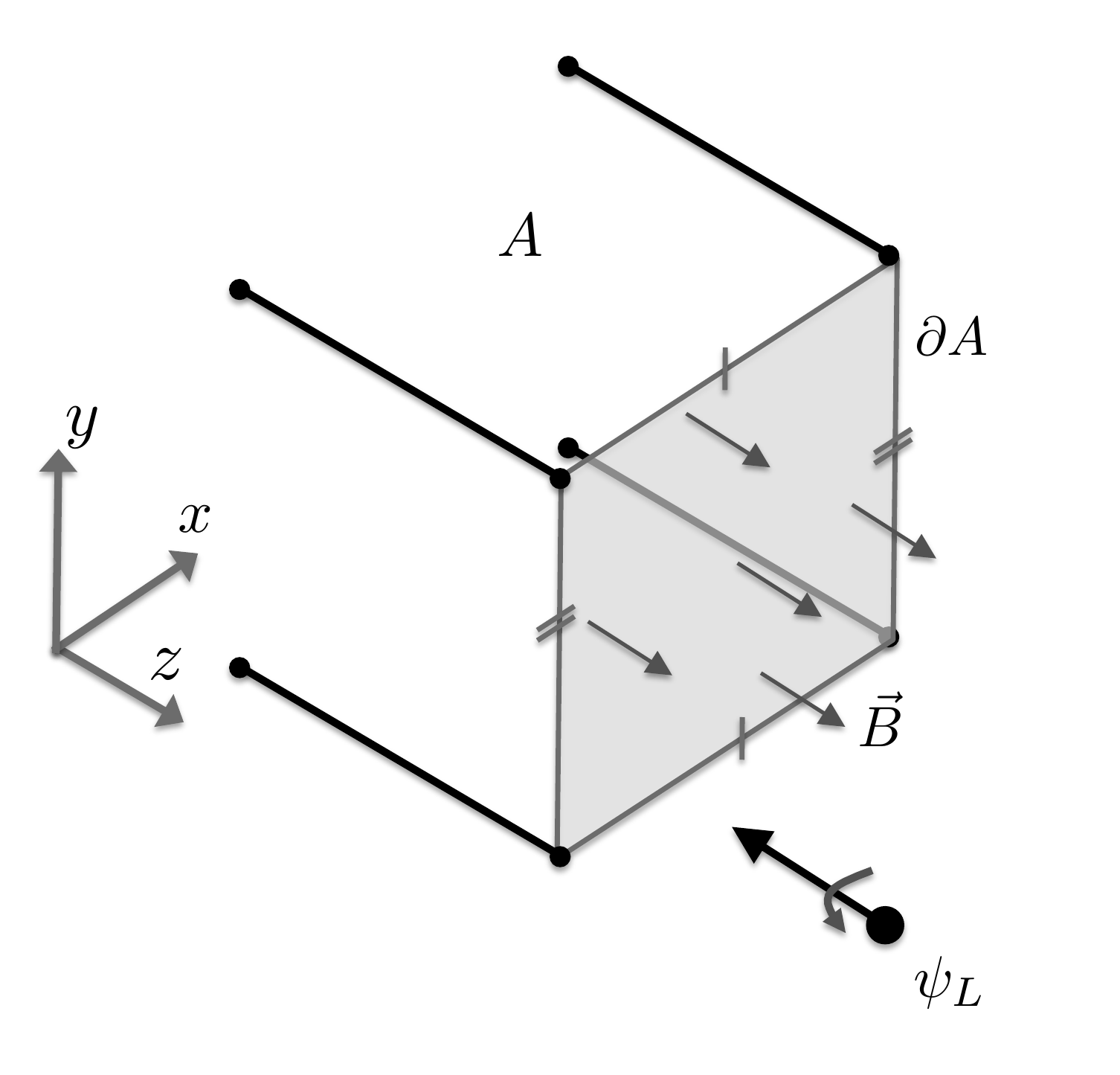}
\end{center}
\vskip -0.5cm
\caption{Weyl fermion studied on $\mathbb{R}^{1,1} \times T^2$ with magnetic flux on $T^2$. Lowest energy modes have spin aligned with magnetic field, and chiral nature of Weyl fermion means that velocity is anti-aligned with spin: thus the low energy physics is that of a chiral CFT$_2$. We study the entanglement entropy of a region $A$ that is a product of an interval on $\mathbb{R}^{1,1}$ and $T^2$.}\label{fig:kktorus}
\end{figure}

Consider now this theory on a spacetime of the form $\mathbb{R}^{1,1} \times T^2$, with $x,y$ parametrizing the torus and $t,z$ parametrizing $\mathbb{R}^{1,1}$. We put a background magnetic flux $\Phi = \int F_{xy}\,dx\,dy$ on the torus, so that the magnetic field points in the $z$ direction, as shown in Figure \ref{fig:kktorus}.

We are now interested in the low-energy physics of this system in the $(t,z)$ directions. We work in the gauge $A_{x} = B y$. Our spinor conventions are in Appendix \ref{app:fermions}: the Weyl equation for the left-handed spinor is
\be
\le[\p_t - \sig^3 \p_{z} + D_{\perp}\ri]\psi_L = 0 \qquad D_{\perp} \equiv -\sig^1(\p_{x} - i q B_y) - \sig^2\p_y \label{weyleq}
\ee
We now search for zero-energy eigenspinors of the Dirac operator on the torus $D_{\perp}$. These take the form
\be
D_{\perp}\chi_p(x,y) = 0 \qquad \chi^+_p = e^{ipx}\le(\begin{tabular}{c}$e^{-\ha Bq\le(y - \frac{p}{Bq}\ri)^2}$ \\ $0$ \end{tabular}\ri),Bq > 0 \qquad \chi_p^- = e^{ipx}\le(\begin{tabular}{c} $0$ \\ $e^{\ha Bq\le(y - \frac{p}{Bq}\ri)^2}$ \end{tabular}\ri),Bq < 0 \label{zerom}
\ee
These modes have zero energy due to a cancellation between the positive zero-point cyclotron energy of the fermion in the lowest Landau level and the Zeeman coupling $B \cdot S$ between the spin of the fermion and the background magnetic field; it is clear that the spin of the zero-mode is anti-correlated with the sign of $Bq$. As usual, their degeneracy can be understood heuristically\footnote{As in most textbook computations of Landau levels, we have been quick; the wavefunctions exhibited do not precisely satisfy torus boundary conditions. These wavefunctions should be understood as being approximately correct in the limit that $B \gg L_{x,y}^{-2}$, and our counting of the degeneracy of the levels is technically only correct in this limit, although an index theorem actually guarantees that it is precisely correct. The exact wavefunctions on the torus can be found in e.g. \cite{fradkin2013field}.}  by noting that the momentum $p$ in the $x$ direction is quantized in units of $\frac{2\pi}{L_x}$ and furthermore that it must be bounded by $\frac{B q L_y}{2}$ to ensure that the center of the wavefunction remains inside the torus, meaning that the number of zero modes is $N_0 = \frac{\Phi}{2\pi}$.

Inserting these wavefunctions into the Weyl equation \eqref{weyleq} we see that lowest-lying modes obey the equation
\be
\psi_L = \chi^{\pm}(x,y)\Psi^{\pm}(z,t) \qquad \le(\p_t - \p_z\ri)\Psi^+(z,t) = 0 \qquad \le(\p_t + \p_z\ri)\Psi^-(z,t) = 0
\ee
As we use $\chi^+$ for positive $Bq$ and $\chi^{-}$ for negative $Bq$, we see that the low-lying mode always propagates chirally along the direction of the magnetic field. Physically, this arises because the 4d Weyl fermion has a definite helicity, meaning that its direction of motion is correlated with the spin, which is correlated with the field. Note that the right-handed antiparticle has the opposite charge but is anti-aligned with the spin, which means that it propagates in the same direction as the particle.

In other words, the low energy dynamics of this system is described by a 2d CFT with
\be
c_L - c_R = \frac{q\Phi}{2\pi}
\ee
This result holds much more generally than the derivation we just gave. We may replace the compact $T^2$ with any compact 2d manifold $\sM_2$ with a flux $\Phi$ through it. The index theorem for the 2d Dirac operator tells us that the number of definite chirality zero modes on $\sM_2$ satisfies $N^+ - N^- = \frac{q}{2\pi}\int_{\sM_2} F$. Four-dimensional chirality is a product of chirality on $\mathbb{R}^{1,1}$ and chirality on $\sM_2$, so a four-dimensional Weyl spinor decomposes as:
\be
\Psi^{4d}_R = \Psi^{\mathbb{R}^{1,1}}_L \otimes \Psi^{\sM_2}_L + \Psi^{\mathbb{R}^{1,1}}_R \otimes \Psi^{\sM_2}_R \ .
\ee
Thus each zero mode of definite chirality on $\sM_2$ gives us a definite chirality spinor on $\mathbb{R}^{1,1}$.

As we now have an effective 2d theory with $c_L \neq c_R$, we expect the physics of Section \ref{sec:2d} to apply. In particular if we consider computing the entanglement of a region that is a product of $\sM_2$ and an interval in $z$, we expect a net entanglement anomaly of the form \eqref{Sans}
\be
\delta_\xi S = \frac{q(N_L - N_R)}{24\pi}\le(\int_{\sM_2} F\ri)\le(\delta\chi_1 + \delta\chi_2\ri), \label{effchiral2d}
\ee
where the boost in question here is in the $z$ direction and does not depend on the compact directions, and we now allow $N_L$ and $N_R$ species of left and right-handed Weyl fermions. In the next section we will demonstrate that this expression can be obtained from a local integral over the entangling surface.

\subsection{Path integral derivation of diffeomorphism anomaly}\label{sec:4dpathint}
We now turn to a derivation of a formula for the variation of the entanglement entropy under an infinitesimal diffeomorphism in a general theory. This precisely parallels the discussion of the two-dimensional case in Section \ref{sec:2dpathint}, and so we only highlight the differences. The variation of the partition function under an infinitesimal diffeomorphism is
\be
\delta_{\xi}\ln \overline{Z}[a,g] = -\int_{\sM_4} d^4x \sqrt{g}\le(\ha \langle \overline{T}^{\mu\nu}\rangle\delta_{\xi}g_{\mu\nu} + \langle \overline{j}^{\mu}\rangle\delta_{\xi} a_{\mu}\ri)
\ee
where the variation of the sources under the diffeomorphism is the usual Lie derivative
\be
\delta_{\xi}g_{\mu\nu} = \nabla_{\mu}\xi_{\nu} + \nabla_{\nu}\xi_{\mu} \qquad \delta_{\xi}a_{\mu} = \xi^{\sig}\nabla_{\sig}a_{\mu} + \le(\nabla_{\mu}\xi^{\sig}\ri)a_{\sig}
\ee
Integrating by parts and using the Ward identities \eqref{Wardnice} we find
\be
\delta_{\xi}\ln \overline{Z}[a,g] = -\frac{i c_m}{2} \int_{\sM_4} d^4x\;\xi^{\nu}\p_{\lam}\le(\sqrt{g} \ep^{\ka\sig\al\beta} F_{\ka\sig} \p_{\al} \Ga^{\lam}_{\nu\beta}\ri) - \int_{\p \sM_4} d^3x \sqrt{\ga} n_{\mu}\le(\overline{j}^{\mu}(\xi^{\sig}a_{\sig}) + \overline{T}^{\mu\nu}\xi_{\nu}\ri)
\ee
We now need to evaluate this variation on an $n$-sheeted replica manifold. We zoom in near a patch of the entangling surface and use coordinates $(r,\th,x^{a,b})$, where $x^{a,b}$ run along the entangling surface. We work with the same conical regulator metric:
\be
ds^2_{(n),a} = f\le(\frac{r}{a}\ri)^2 dr^2 + r^2 d\th^2 + dx^i dx^j \delta_{ij}, \label{regulated4d}
\ee
but we allow now the diffeomorphism $\xi^{\mu}(x)$ and the background field strength $F_{\mu\nu}$ to be arbitrary functions of $x^a$ and $r$.

We compute first the bulk term, which is
\be
\delta_{\mathrm{bulk}} \equiv -\frac{i c_m}{2}\int_{\sM_4} d^4x\;\xi^{\nu}\p_{\lam}\le(\sqrt{g} \ep^{\ka\sig\al\beta} F_{\ka\sig} \p_{\al} \Ga^{\lam}_{\nu\beta}\ri) = - 2 i c_m \pi n \int dr d^2x^{i} \xi^{\th}(r)\p_r\le(F_{x^{i} x^{j}}(r) I_a(r)\ri)
\ee
with $I_a(r)$ the same as in \eqref{regexp}. As argued there, as we take $a \to 0$, $I_a(r)$ approaches $-1$ almost everywhere, and $\p_r I_a(r)$ approaches a delta function at $r = 0$ with weight $\frac{1}{n^2}-1$. Thus we can expand out the derivative in $r$ to find
\be
\lim_{a \to 0}\delta_{\mathrm{bulk}} = -2 i c_m \pi \int d^2x^{i} \le[\le(\frac{1}{n}-n\ri) \le(F_{x^{i} x^{j}}\ri)\xi^{\th}\bigg|_{r \to 0} -  n \int dr \p_r(F_{x^{i}x^j})\xi^{\th}\ri]
\ee
The first term here is the desired local expression on the entangling surface. The second term is not local, as it essentially measures the change in $F$ from the entangling surface to infinity. However we see that it is linear in $n$: in other words, it is merely renormalizing the partition function and is not associated with entanglement, and will cancel against the denominator in \eqref{renyidef}.

We turn now to the evaluation of the boundary term. Following the reasoning of the earlier section, we know that the full contribution will come from the Bardeen-Zumino term \eqref{4dBZ}. It is easy to see that $\overline{j}^{\mu}_{BZ} = 0$. Computing $\overline{T}^{\mu\nu}_{BZ}$ explicitly we find that the boundary term is
\be
\delta_{\mathrm{bdy}} \equiv  \int_{\p \sM_4} d^3x \sqrt{\ga}\;n_{\mu}\overline{T}_{BZ}^{\mu\nu}\xi_{\nu} = \le(\frac{6 i c_m \pi n}{f\le(\frac{r}{a}\ri)^2}\int d^2 x^{i} F_{x^i x^j}\xi^{\th}\ri)\bigg|_{r \to 0},
\ee
where we have neglected terms that do not contribute in the $r \to 0$ limit. To find the variation of the Renyi entropy we assemble these pieces, taking care to also take into account the denominator of \eqref{renyidef}. We find
\be
\delta_{\xi} S_n = 4 \pi i c_m  \le(1 + \frac{1}{n}\ri) \int d^2x F_{x^i x^j}\xi^{\th}
\ee
This expression may be written more covariantly as
\be
\delta_{\xi} S_n = 2 \pi i c_m\le(1+\frac{1}{n}\ri) \int_{\p A} d\Sig^{\mu\nu} F_{\mu\nu} \le( \nabla_{\rho}\xi_{\sig}\ri)\ep_{\p A}^{\rho\sig}
\ee
where $\ep_{\p A}^{\mu\nu}$ is the binormal to the entangling surface $\p A$, defined in Appendix \ref{app:geom} in \eqref{binormal}. Finally, we rotate to Lorentzian signature by defining a local boost around the entangling surface via $2 \delta\chi = i \ep_{\p A}^{\mu\nu}\nabla_{\mu}\xi_{\nu}$. We also take the $n \to 1$ limit to find the following elegant expression for the entanglement anomaly
\be
\boxed{\delta_{\xi}S = -8\pi c_m \int_{\p A} (F \delta \chi)} \label{ans4dcons}
\ee
This reduces to the expected expression \eqref{effchiral2d} for free fermions if we study the system on the product of a compact 2d manifold with flux and $\mathbb{R}^{1,1}$ and use the known value for the anomaly coefficient $c_m = \frac{q(N_L - N_R)}{192\pi^2}$.

\subsection{Weyl fermions and chiral zero modes with twist flux} \label{sec:twistedzeros}
In the previous section we argued that if we couple Weyl fermions to a non-trivial background $U(1)$ gauge potential, then the presence of zero modes leads us to expect an anomalous transformation of the entanglement entropy under diffeomorphisms. In this section we describe the converse problem: we couple Weyl fermions to a background gravitational curvature, then argue that the presence of zero modes leads us to expect an anomalous transformation of the entanglement entropy under $U(1)$ {\it gauge} transformations. The ``zero'' modes in question are not as familiar as those above, though we expect them to be related to the chiral vortical effect \cite{Landsteiner:2011cp,Son:2009tf}.

Consider the following gravitational background, viewed as an infinitesimal deformation of the product manifold that is Rindler space times $\mathbb{R}^2$:
\be
ds^2 = dr^2 + r^2(-d\eta + U_i(x_i) dx^i)^2 + dx^i dx^j \delta_{ij} + \sO(U^2)
\ee
Here $i,j$ run over the coordinates $x,y$ on the $\mathbb{R}^2$. $U_i$ is an Abelian gauge field that generates translations in $\eta$; these translations correspond to $SO(1,1)$ rotations of the normal frame around the entangling surface, and so we will call $U_i$ the ``twist'' gauge field. This nomenclature makes slightly more sense in Euclidean signature, where the relevant transformations are really $SO(2)$ twists of the normal frame around the entangling surface. We present some geometric background in Appendix \ref{app:geom}.

Now let $U_i$ have a constant field strength, i.e. $dU = \sB dx \wedge dy$. We note that this is somewhat similar to the well-known Lorentzian Taub-NUT solution, which has a similar structure, except that the twist flux of the $U(1)$ gauge field there is finite and distributed over a compact $S^2$ rather than an infinite $\mathbb{R}^2$. In the usual compact Taub-NUT global issues force the time coordinate to be periodically identified in units of the total twist flux (also called the ``NUT charge''). Here the total twist flux on the $\mathbb{R}^2$ is formally infinite, and we will simply ignore any global issues.

We now work out the equation of motion for a Weyl fermion living on this background. This computation is standard, and so we relegate all details of the derivation to Appendix \ref{app:fermions} and write down only the final equation of motion for the two-component left-handed Weyl spinor $\psi_L$:
\be
\le(\sig^{z}\le(r \p_r + \ha \ri) - \p_{\eta} + r D_{\perp}\ri)\psi_L = 0
\ee
where the operator acting on the transverse space is
\be
D_{\perp} \equiv \sig^i\le(\p_i + U_i \p_{\eta}\ri), \label{DperpNUT}
\ee
and where we are working close to the Rindler horizon $r \to 0$, i.e. we have neglected terms of the form $r^2 \sB$. The transverse operator \eqref{DperpNUT} couples the fermion to $U_i$ as though it was a $U(1)$ gauge field, except that the $U(1)$ charge is the Rindler energy. We now specialize to the time dependence for all fields of the form $e^{-i\om\eta}$ and the gauge $U_x = \sB y$; the transverse operator becomes
\be
D_{\perp} = \sig^x\le(\p_x - i \om \sB y\ri) + \sig^y \p_y,
\ee
i.e. precisely equivalent to \eqref{weyleq} in the previous section with the important substitution of the Rindler energy $\om$ for the the $U(1)$ charge $q$. Thus the zero modes $D_{\perp} \chi_{p}^{\pm}(x,y) = 0$ take precisely same form as in \eqref{zerom}, except that their 2d chirality is now determined by the sign of $\om\sB$: interestingly, positive frequency modes have the opposite chirality to negative frequency ones. Building a full spinor from $\psi_L = \chi^{\pm}(x,y)\Psi_{\om}^{\pm}(r)e^{-i\om\eta}$, we find for the equation of motion in the Rindler radial coordinate:
\begin{align}
\le(r\p_r + \ha + i \om\ri)\Psi_{\om}^{+}(r) & = 0,\qquad \om\sB > 0 \\
\le(r\p_r + \ha - i \om\ri)\Psi_{\om}^{-}(r) & = 0,\qquad \om\sB < 0
\end{align}
For concreteness, let us now fix $\sB > 0$. Then we see that independently of the sign of $\om$, we always have
\be
\Psi^{\pm}_{\om}(r) \sim \frac{1}{\sqrt{r}} e^{-i|\om| \log r}
\ee
In other words, both positive and negative frequency modes always have the same definite sign for their spatial momentum in the $r$-direction.

This is somewhat novel. Normally one interprets negative frequency modes as anti-particles, meaning that one should invert all their quantum numbers when considering a physical excitation. This inversion means that the anti-particles always move in the opposite direction from the particles, and thus that the net {\it charge} current in the radial direction should now have a definite sign. Note that the degeneracy of these chirally propagating charge modes is given by $|\om| \int d^2x \sB$, i.e. it depends on their frequency.

Thus the current receives a contribution only from the zero modes. The net flow of current through the Rindler horizon at Rindler temperature $\beta$ is computed by summing over these modes:
\be
\langle j^r(r) \rangle = \frac{2q}{r} \frac{{\sB} A}{2\pi} \int_0^{\infty} \frac{d\om}{2\pi} \frac{\om}{1 + e^{\beta\om}} = \frac{q}{24} \frac{{\sB} A}{\beta^2 r} \ . \label{lorj}
\ee
This expression is derived in Appendix \ref{app:fermions}. The existence of a net charge flow off the edge of the system means that it is no longer gauge-invariant. To turn this fact into a precise statement about the entanglement entropy, consider computing the entanglement entropy of the Rindler wedge $r>0$ from the Euclidean partition function as
\be
S = \le(1 - \beta\frac{\p}{\p\beta}\ri)\log Z(\beta)\bigg|_{\beta = 2\pi} \label{EEpart}
\ee
with $\beta$ the Rindler temperature. Now consider the gauge variation of the system with gauge parameter $\Lam(r)$, excluding a small disc near the region $r = 0$. The partition function changes as
\be
\delta_{\Lam} \log Z(\beta) = -\int_{\sM} d^4x \langle j^{\mu}\rangle \p_{\mu}\Lam = -\int_{\p\sM} d^3x \le(n_{\mu} \langle j^{\mu} \rangle \ri) \Lam(x),
\ee
where we have integrated by parts. The anomaly plays no direct role here as the geometry away from the tip of the cone is trivial. The second term is simply the net flow of current through the surface at $r = 0$. In components this is
\be
\delta_{\Lam} \log Z = \lim_{r \to 0}\int d^2x\;r\beta \langle j^r \rangle \Lam(r)
\ee
where we have performed the integral over Euclidean time with period $\beta$. We now use the Lorentzian expression for $j^r$ from \eqref{lorj} and plug into \eqref{EEpart} to conclude that
\be
\delta_{\Lam}S = (N_L-N_R)\frac{q \sB A}{24\pi}\Lam(r \to 0) \ , \label{gaugevar}
\ee
with $A$ the transverse area, and where we have again generalized to arbitrary numbers of left and right-handed fermion species. In other words, in the presence of a nonzero twist flux, the entanglement entropy is sensitive to $U(1)$ gauge transformations with support on the entangling surface. This formula is clearly the gauge analog of \eqref{effchiral2d}.

\subsection{Path integral derivation of gauge anomaly} \label{sec:gaugeanomaly}
We now perform a similar path integral argument to understand the {\it gauge} variation of the entanglement entropy. As we are now working on a non-trivial gravitational background, we work in the diff-preserving frame \eqref{Wardmixedorig}.  Under a $U(1)$ gauge transformation with compact support, the transformation of the partition function is
\be
\delta_{\Lam} \log Z[a,g] = \int d^4x \sqrt{g}\;\p_{\mu} \langle j^{\mu} \rangle \Lam(x) \ .
\ee
From the anomaly equation \eqref{Wardmixedorig}, this can be written as
\be
\delta_{\Lam} \log Z[a,g] = -i\frac{c_m}{4}\int d^4x \sqrt{g}\;\Lam(x) \ep^{\ka\sig\al\beta} R^{\nu}_{\phantom{\nu}\lam\ka\sig} R^{\lam}_{\phantom{\lam}\nu\al\beta} \label{gaugevar1}
\ee
Thus to determine the variation of the entanglement entropy, we need to evaluate the geometric invariant in \eqref{gaugevar1} on the $n$-sheeted geometry and then extract the dependence on $n$ as we take $n \to 1$. A general formula for such expressions was found (in the context of evaluating holographic entanglement entropy in higher-derivative theories of gravity) in \cite{Dong:2013qoa,Camps:2013zua}. Applying the prescription of \cite{Dong:2013qoa} to the expression above, we find after some algebra that
\be
\delta_{\Lam}S_{EE} =  16\pi \frac{ic_m}{4} \int_{\p A} d\Sig^{\ga\delta} \ep^{\mu\rho}_{\p A} \le(R_{\mu\rho\ga\delta} + 2 K_{\mu\al\ga} K_{\rho\beta\delta} g^{\al\beta}\ri)\Lam(x)
\ee
where $K_{\mu\al\beta}$ is the extrinsic curvature. With the help of the Voss-Ricci equation -- described in the Appendix in \eqref{vossricci} -- we see that the combination of extrinsic curvatures and Riemann tensor that appears here actually measures the field strength $\Om$ of the twist gauge field $V$, as defined in \eqref{simpomega}:
\be
\delta_{\Lam}S_{EE} =  8\pi i c_m \int_{\p A} d\Sig^{\ga\delta} \Om_{\ga\delta}\Lam(x)
\ee
Finally we should analytically continue to Lorentzian signature. The Euclidean gauge field $V_{\mu}$ that generates $SO(2)$ rotations in the normal bundle is continued to the Lorentzian signature gauge field $U_{\mu}$ of Section \ref{sec:twistedzeros} via $V_{\mu} = i U_{\mu}$. Thus we find finally
\be
\boxed{\delta_{\Lam}S_{EE} =  8\pi c_m \int_{\p A} (dU \Lam)} \label{gengauge}
\ee
This is the gauge analog of \eqref{ans4dcons}, and it again reproduces the free fermion computation \eqref{gaugevar} if we use the relation $c_m = \frac{q(N_L - N_R)}{192\pi^2}$ for free Weyl fermions.

\subsection{The choice of anomaly frame} \label{sec:anomalyframe}

An important point is that our generating functional arguments agree with the free field analysis only when we are working in the correct anomaly frame: i.e. the free field diffeomorphism anomaly \eqref{effchiral2d} agrees with the general formula \eqref{ans4dcons} only in the gauge-preserving frame, whereas the free field gauge anomaly \eqref{gaugevar} agrees with the general formula \eqref{gengauge} only in the diff-preserving frame. We believe this is because we need to turn on background fields to {\it see} each of these anomalies, and this can only be safely done in the appropriate anomaly frame. 

More concretely, the background fields in question grow linearly with space (e.g. $A_x \sim B y$). For the magnetic field case we imagined compactifying on a torus; to make the gauge field compatible with torus boundary conditions we need to perform a large gauge transformation, which is problematic if there is an anomaly. We could imagine simply taking an infinite $\mathbb{R}^2$ rather than a $T^2$ (this was indeed always the case for the twist flux, where we do not have a useful notion of twist flux quanitzation). However now the fields grow arbitrarily large as we move outwards, and there may actually be boundary terms from infinity that scale geometrically the same way as the volume, affecting the answer. The safest way to perform the computation when turning on background fields seems to be to work in the anomaly frame that is appropriate to the background field in question. 

Another way to say this is that adding the anomaly-shifting counterterm \eqref{counterterm} changes the free-field computation by adding a term to the current through the entangling surface \eqref{lorj}: the full contribution to the answer comes from the free modes only in the appropriate anomaly frame.

\section{Mixed anomaly on the boundary of a 5d Hall phase} \label{sec:4dpathintbdy}

We now study the 4d theory described above as the boundary of a gapped 5d ``Hall'' phase. We begin with the original form of the Ward identities in \eqref{Wardmixedorig}, {\it without} the addition of the anomaly-shifting counterterm \eqref{counterterm}. We would now like to supplement this system with a five-dimensional gapped theory which is the analog of the ``Hall droplet'' discussed above. The full generating functional is then
\be
W_{tot}[A,G] = W[a,g] + i c_m S_{CS}[A,G],
\ee
where $A$ and $G$ live in five dimensions and the appropriate five-dimensional Chern-Simons term is
\be
S_{CS}[A,G] \equiv \frac{1}{4}\int_{\sM_5} d^5x\;\bar{\ep}^{PQMNR} A_P \sR^{A}_{\phantom{A}BQM} \sR^{B}_{\phantom{B}ANR} \ .
\ee
One may check that under both a 5d $U(1)$ gauge transformation $\Lam$ and a 5d diffeomorphism $\chi$ we have
\be
\delta_{\Lam}W_{tot}[G,A] = 0 \qquad \delta_{\chi} W_{tot}[G,A] = 0 \ .
\ee
Furthermore, under a general variation of background fields we have (in direct analogy to \eqref{totvar})
\begin{align}
\delta W_{tot}[G,A] & = \int_{\p \sM_5} d^4x \sqrt{g}\le[ \le(j^{\mu}\ri)\delta a_{\mu} + \ha\le(T^{\mu\nu} + T^{\mu\nu}_{BZ}\ri)\delta g_{\mu\nu}\ri] \nonumber \\ & + i c_m \int_{\sM_5} \sqrt{G} \le(\sJ^M  \delta A_M + \sC^{MN} \delta G_{MN}\ri)
\end{align}
We refer the reader back to \eqref{totvar} for an explanation of where these terms come from. Just as in the lower-dimensional case, the covariant stress tensor and current are now defined as $j^{\mu}_{cov} = j^{\mu}$, $T^{\mu\nu}_{cov} = T^{\mu\nu} + T^{\mu\nu}_{BZ}$ (in this presentation of the anomaly the Bardeen-Zumino correction term for the current vanishes: $j^{\mu}_{BZ} = 0$) and their (non)-conservation equations are as follows:
\begin{align}
\nabla_{\mu}j_{cov}^{\mu} & =  i\frac{c_m}{4} \ep^{\kappa\sig\al\beta} R^{\nu}_{\phantom{\nu}\lam\ka\sig} R^{\lam}_{\phantom{\lam}\al\beta} \nonumber \\
\nabla_{\mu}T_{cov}^{\mu\nu} & = f^{\mu}_{\phantom{\mu}\nu} j^{\nu} + i\frac{c_m}{2} \nabla_{\nu}\le(\ep^{\rho\sig\al\beta} F_{\rho\sig} R^{\mu\nu}_{\phantom{\mu\nu}\al\beta}\ri) \label{covanomT4d}
 \end{align}
where we have repeated the Ward identity for $j^{\mu} = j^{\mu}_{cov}$ purely for convenience. We note that if we had started with the presentation of the anomaly {\it with} the anomaly-shifting counterterm \eqref{counterterm}, the definitions of both the original currents and the Bardeen-Zumino correction terms would differ, but the final covariant currents would be the same, i.e. $j^{\mu} + j^{\mu}_{BZ} = \overline{j}^{\mu} + \overline{j}^{\mu}_{BZ} = j^{\mu}_{cov}$ and similarly for $T^{\mu\nu}$.

We now seek to understand how the entanglement entropy transforms under a 4d diffeomorphism and gauge symmetry, and thus we seek to understand the analog of the ``sheared metric'' \eqref{shearmet} in the four-dimensional case. The metric part of this takes precisely the same form:
\be
ds^2 \equiv G_{MN}^{(\xi)} dX^M dX^N = \le[g_{\mu\nu} + f(z)\le(\nabla_{\mu} \xi_{\nu} + \nabla_{\nu} \xi_{\mu}\ri)\ri] dx^{\mu} dx^{\nu} + dz^2 \label{shearmet2}
\ee
where as before $f(z)$ interpolates between zero at $z = \infty$ and $1$ at $z = 0$. We will also now allow a ``shear'' of the gauge field, which we take to be:
\be
A^{(\xi)} = a + f(z) d\Lam \qquad  \label{gaugedef}
\ee
In other words, for the 5d metric we interpolate between two 4d metrics related by a 4d diffeomorphism $\xi$. For the 5d gauge field we interpolate between two 4d gauge fields related by a 4d gauge transformation $\Lam$. For notational convenience we will simply refer to this total operation as $\overline{\delta}_{\xi}$.

We now study the entanglement entropy by studying the partition function on the $n$-sheeted replica geometry and performing the variation $\overline{\delta}_{\xi}$. The boundary geometry and extension into the bulk is precisely as discussed in Section \ref{sec:2dpathintbdy}, except that everything is now in higher dimension. The analog of \eqref{varxi} is
\be
\overline{\delta}_{\xi}W_{tot}[A,G_{(n)}] = \int_{\p\sM_5}d^4x\le(\ha T^{\mu\nu}_{cov} \overline{\delta}_{\xi} g_{\mu\nu} + j^{\mu}_{cov}\overline{\delta}_{\xi} a_{\mu}\ri) + i c_m \int_{\sM_5} d^5x \sqrt{G_{(n)}} \le(\sJ^M  \overline{\delta}_{\xi} A_M + \sC^{MN} \overline{\delta}_{\xi} G_{MN}\ri) \label{compvar}
\ee
The bulk 5d terms $\sJ^M$ and $\sC^{MN}$ are found to be (see e.g. \cite{Landsteiner:2011iq} for a derivation):
\begin{align}
\sJ^P & = \frac{1}{4} \bar{\ep}^{PQMNR} A_P \sR^{A}_{\phantom{A}BQM} \sR^{B}_{\phantom{B}ANR} \\
\sC^{MN} & = \frac{1}{4}\le(\nabla_T R^{TM}_{\phantom{TM}PQ} F_{RS} \bar{\ep}^{NPQRS}\ri) \ .
\end{align}
Paralleling our discussion of the Cotton tensor in the earlier section, they can be shown to have no contribution. We turn now to the first term in \eqref{compvar}. The variation of the boundary metric and gauge field are those of a combined diffeomorphism and gauge transformation:
\be
\overline{\delta}_{\xi}g_{\mu\nu} = \nabla_{\mu}\xi_{\nu} + \nabla_{\nu}\xi_{\mu} \qquad \overline{\delta}_{\xi}a_{\mu} = \xi^{\sig}\nabla_{\sig}a_{\mu} + \le(\nabla_{\mu}\xi^{\sig}\ri)a_{\sig} + \p_{\mu}\Lam
\ee
Integrating by parts on the boundary terms we find
\be
\overline{\delta}_{\xi}W_{tot}[A,G_{(n)}] = \int_{\p\sM_5}d^4x\sqrt{g}\le[\le(-\nabla_{\mu} T^{\mu\nu}_{cov} + f^{\nu}_{\phantom{\nu}\mu}j^{\mu}_{cov}\ri) \xi_{\nu} - \le(\nabla_{\mu} j^{\mu}_{cov}\ri)\le(\xi^{\sig}a_{\sig} + \Lambda\ri)\ri] \label{totvar5d}
\ee
The second term is absent in a non-anomalous theory where the current is conserved. We note that if we choose to pick $\Lam$ to be $\Lam \equiv -\xi^{\mu} a_{\mu}$ we may cancel it. However the first term cannot be canceled: using the expression for the covariant anomaly \eqref{covanomT4d} we find for the contribution from the first term:
\be
\overline{\delta}_{\xi}W_{tot}[A,G_{(n)}]_{(1)} = i\frac{c_m}{2} \int d^4x \sqrt{g}\;\ep^{\ka\sig\al\beta}F_{\ka\sig} R^{\mu\nu}_{\phantom{\mu\nu}\al\beta}\le(\nabla_\nu\xi_{\mu}\ri) \label{Wtrans4d}
\ee
Now on a conical deficit with opening angle $2\pi(n-1)$ the Riemann tensor can be understood as having a delta function singularity on the (two-dimensional) worldvolume $\p A$ of the deficit \cite{Fursaev:1995ef}:
\be
R^{\al\nu}_{\phantom{\al\nu}\rho\sig}(x) = 2\pi(n-1) \ep_{\p A}^{\al\nu} \ep^{\p A}_{\rho\sig} \delta_{\p A}(x),
\ee
where $\ep_{\p A}^{\al\nu}$ is the binormal defined in \eqref{binormal}. Putting this into \eqref{Wtrans4d} and extracting the linear dependence on the opening angle, we find for the anomalous variation of the entanglement entropy arising from the first term:
\be
\overline{\delta}_{\xi} S_{EE,(1)} = 4\pi i c_m \int_{\p A} \le(d\Sig^{\mu\nu} F_{\mu\nu}\ri) \ep_{\p A}^{\al\beta} \nabla_{\al}\xi_{\beta}
\ee
with $d\Sig_{\mu\nu}$ the area element.

We now turn to the second term in \eqref{totvar5d}, that involving the non-conservation of the current. As the form of the covariant anomaly \eqref{covanomT4d} for the divergence of the current takes precisely the same as the consistent form exhibited in \eqref{Wardmixedorig}, its contribution may be found following precisely the same logic as in Section \ref{sec:gaugeanomaly}, where the gauge parameter there is now replaced by the combination $\le(\xi^{\sig}a_{\sig} + \Lambda\ri)$. We find for the contribution from the second term:
\be
\overline{\delta}_{\xi}S_{EE,(2)} = 8\pi c_m \int_{\p A} dU (\Lam + \xi^{\sig} a_{\sig}),
\ee
with $U$ the twist gauge field. Defining a local boost as before by $2 \delta\chi = i \ep_{\p A}^{\mu\nu}\nabla_{\mu}\xi_{\nu}$, and assembling the two pieces, we find for the variation of the entanglement entropy:
\be
\boxed{\overline{\delta}_{\xi} S_{EE} = 8 \pi c_m \le(\int_{\p A} (-F \delta \chi) + dU (\Lam + \xi^{\sig} a_{\sig})\ri)},
\ee
We note that unlike the expressions discussed in the previous section, there is no notion of a gauge or diff-preserving frame and indeed no ambiguity in anomaly frame at all. The covariant anomaly takes a unique form, and we thus have a unique answer for the transformation of the entanglement entropy in this construction.

\section{Gravitational and mixed anomalies in six dimensions} \label{sec:6dindex}

In six dimensions we restrict ourselves to a free-field analysis. Here there are two different kinds of chiral fields that one can study: free Weyl fermions and the self-dual boson.  The helicity group is $Spin(4) = SU(2)_L \times SU(2)_R$, and we will define a right-handed particle as one which has nonzero $SU(2)_R$ helicity.  Using this definition, we find that the entanglement anomaly is $-8$ times larger for the right-handed (or self-dual) boson than for the right-handed fermion, as expected from their contributions to the gravitational anomaly.

{\bf Weyl fermion.} By analogy to previous sections, we will consider a six-dimensional Weyl fermion on a product manifold $\mathbb{R}^{1,1} \times \sM_4$, where $\sM_4$ is a compact Euclidean four-manifold. We are interested in the spectrum of massless modes on the non-compact $\mathbb{R}^{1,1}$ component.  If there are more right-moving modes than left-moving ones, then we expect there to be an entanglement anomaly.  For a 6d right-handed Weyl fermion, accroding to our helicity conventions, a left-chirality mode on $\sM_4$ will be left-moving on the $\mathbb{R}^{1,1}$ factor, while a right-chirality mode will be right-moving:
\be\label{split}
\Psi^{6d}_R = \Psi^{4d}_R \otimes \Psi_R^{2d} + \Psi^{4d}_L \otimes \Psi^{2d}_L + \text{massive 2d modes}.
\ee
where the massive $2d$ modes move in both directions, and hence do not contribute to the anomaly.

Now the spectrum of the Dirac operator on $\sM_4$ is constrained by an index theorem, which tells us that the difference in the number of (complex) left-chirality modes and right-chirality modes is given by a topological invariant of $\sM_4$:
\be
n_R - n_L = -\frac{P}{24}
\ee
with $P$ the Pontryagin number $P = \frac{1}{8\pi^2}\int_{\sM_4}\tr( R \wedge R)$ \cite{ChristensenDuff79}.  Because the spinor representations of Spin(4) are pseudoreal, $n_L$ and $n_R$ are both constrained to be \emph{even}.  Note that for a smooth four-dimensional manifold with spin structure, $P$ is a multiple of 48 by Rokhlin's theorem, as is required for the above relation to make sense.

The above mismatch of zero modes implies an effective chiral 2d CFT on the $\mathbb{R}^{1,1}$ with $c_R - c_L = -\frac{P}{24}$ (where $c_{L/R} = n_{L/R}$ because a Weyl spinor is complex).  Hence the corresponding entanglement anomaly for a single Weyl fermion can be obtained from the 2d result \eqref{Sans}:
\be
\delta S = -\frac{1}{12}\frac{P}{24}\delta \chi
\ee
This can be obtained from a local expression on the 4d entangling surface.  Generalizing to the case where there are $(N_L, N_R)$ 6d Weyl fermions, we obtain:
\be
\delta S = -\delta \chi\, \frac{N_R - N_L}{2304 \pi^2} \int_{E} \tr(R \wedge R)  \label{6dboost}
\ee
where $R$ is the intrinsic four-dimensional curvature on the entangling surface.

We may also consider the case in which $\sM_4$ depends nontrivially on the $\mathbb{R}^{1,1}$ directions.  Since the number of left and right moving modes is always an integer, deforming the geometry to allow for extrinsic curvature $K^{ab}$ cannot affect the number of modes, as evaluated on a compact surface, and we expect \eqref{6dboost} to still apply. 

%

{\bf Chiral boson.} In dimensions of the form $D = 4k + 2$, there also exists a bosonic chiral theory, namely a $k/2$-form field with self-dual (i.e. right-handed) field strength.  For $k = 0$, this is just a right-moving chiral scalar field $\Phi_R$, which can be treated using the methods of section \ref{sec:2d}.  For $k = 1$, we have a 2-form potential $A_{ab}$ whose curvature $F = dA$ satisfies the self-duality condition $F = *F$.  The self-duality equation, together with the Bianchi identity $dF = 0$, implies the Maxwell equation $d *F = 0$, and is thus sufficient to determine the equations of motion for the system.  The gauge transformation is $\delta A = d\alpha$, where $\alpha$ is a 1-form.

If we gauge fix using the Lorenz gauge $d * A = 0$, then all solutions to the equations of motion are zero vectors of the 6d Hodge Laplacian:
\be
\Delta A = (d * d *  + * d * d) A = 0
\ee
On any product manfold, the Hodge Laplacian has the property that it decomposes into the sum of the Laplacian of the base and the fiber: $\Delta^{6d} = \Delta^{4d} + \Delta^{2d}$.  The 2-form A splits into 3 polarization classes: a) 2-forms on $\sM_4$ times scalars on the $\mathbb{R}^{1,1}$, b) 2-forms on $\mathbb{R}^{1,1}$ times scalars on $\sM_4$, and c) mixed products of 1-forms on both $\sM_4$ and $\mathbb{R}^{1,1}$.  We will use $u,v$ to represent right-moving and left-moving coordinates respectively on $\mathbb{R}^{1,1}$, and $i,j\ldots$ to represent coordinates of $\sM_4$.

We are interested in finding chiral modes of $\mathbb{R}^{1,1}$ which propagate in only one direction.  Let us begin by restricting attention to modes whose profile on $\mathbb{R}^{1,1}$ is right-moving: $e^{i\omega u}$.  Such modes are massless excitations with $\Delta^{2d} = 0$, and therefore also $\Delta^{4d} = 0$.  The 0, 1, or 2-forms on $\sM_4$ are therefore harmonic, i.e. both $dA$ and $d * A$ vanish on $\sM_4$.  However, the harmonic 0-form solution is pure gauge, and the harmonic 1-form fails to satisfy the self-duality constraint (since $F_{iuv} \ne 0$, but $F_{ijk}|_{\sM_4} = 0$ for a harmonic mode).  It follows that we can restrict attention to the case in which the 2-form is polarized along $\sM_4$.  We are left with the solution
\be
A_{ij} = e^{i\omega v} h_{ij}
\ee
where $h_{ij}$ is a harmonic 2-form on $\sM_4$.  This solution has a field strength polarized in the $h_{ij,v}$ direction, and in order for this field strength to be self-dual, $h_{ij}$ must also be self-dual, since $^{6d}\epsilon^{abu}_{\phantom{abu}cdu} = {}^{4d}\epsilon^{ab}_{\phantom{ab}cd}$.  On the other hand, if we instead chose the field to move in the $u$ direction, then since $^{2d}\epsilon^{v}_{\phantom{v}v} = - {}^{2d}\epsilon^{u}_{\phantom{u}u}$, we would instead need $h_{ij}$ to be anti-self-dual.  (The situation is reversed for a 6d anti-self-dual field.)  These definitions of $\epsilon$ are in accordance with our helicity convention, that
\be
A^{6d}_R = A^{4d}_R \otimes \Phi_R^{2d} + A^{4d}_L \otimes \Phi^{2d}_L + \text{massive 2d modes},
\ee
similiar to the fermion decomposition \eqref{split}.

We conclude that the 6d self-duality constraint is equivalent to saying that the left-chirality modes on $\sM_4$ move left on $\mathbb{R}^{1,1}$, while the right-chirality modes move right---just as in the case of the fermion.  But there are a different number of right and left chirality modes on $\sM_4$, if it has nonzero Pontryagin number.  According to the index theorem:
\be
n_L - n_R = \frac{P}{3},
\ee
where $P$ is always a multiple of 3 on any four-dimensional manifold (without assuming the existence of a spin structure).  We conclude that a self-dual 2-form field has an entanglement anomaly which is given by the same expression \eqref{6dboost} but with a coefficient that is $-8$ times the value for a Weyl fermion.

{\bf General entanglement anomaly formula}. Given these results we may generalize to a less symmetric situation: i.e. consider a product manifold of the form described above, but consider a boost $\delta \chi$ that is nonconstant along the entangling surface.  Since $\delta S$ must be given by an expression local on $E$, in this case we expect \eqref{6dboost} to be valid up to a total derivative term:
\be
\delta S = -16\pi c_1 \int_{E} \le(\tr(R \wedge R ) + D_a v^a\ri) \delta \chi(x), \label{total}
\ee
where we have chosen to define a coefficient $c_1$ such that for fermions and bosons respectively we have
\be
c_{1, \mathrm{fermions}} = -\frac{1}{36864 \pi^3}(N_R - N_L) \qquad c_{1,\mathrm{bosons}} = \frac{8}{36864\pi^3} (N_R - N_L)
\ee
There seems to be one allowed covariant total derivative term\footnote{The only term which is weight 4, parity odd on $\sM_4$, and even on $\mathbb{R}^{1,1}$, is $\ep^{abce} D_a (K_{bc} D_e K_{fh}) g^{fh}$, where $D_a$ is the intrinsic covariant derivative on $\sM_4$. Note that it explicitly involves the inverse metric, and thus has a different weight under homogenous Weyl transformations than the other entanglement anomaly terms that we have studied.} 
, but as we explain below it does not actually appear, i.e. $v^a = 0$ above.

This is as far as we can go from a free-field theory analysis. To complete the story one should now repeat the path-integral derivations presented above for the 6d case to derive this expression from a Ward-identity analysis, which would also hold in less symmetric situations. We have not done that in this paper. 

However, the desired expression was obtained from a holographic analysis in \cite{Loga}, where it was also argued that the resulting expression should hold for any anomalous field theory. If we restrict the result from \cite{Loga} to a product manifold, we do indeed find an expression that is precisely \eqref{total} with no extra total derivative term, where $c_1$ is the anomaly coefficient for one of the two possible 6d gravitational anomalies. 

Interestingly, however, for a non-product manifolds \cite{Loga} finds an extra term in the entanglement anomaly that is sensitive to the integral of the square of the twist flux over the entangling surface, i.e. in our notation there is a contribution of the form
\be
\delta S \sim c \int_{\sM^4} dU \wedge dU \ . \label{twistsquare}
\ee
This coefficient of this term depends on both of the 6d purely gravitational anomaly coefficients.  

To understand this sort of term from free field theory one has to perform an analysis of the type performed in \ref{sec:twistedzeros} in six dimensions on an entangling surface that has a nonzero square of the twist field flux. 

This seems possible (and indeed, as we discuss briefly in Appendix \ref{app:fermions} the degeneracy of zero modes is even consistent with this), but there is reason to expect the existence of subtleties: recall from Section \ref{sec:twistedzeros} that we obtained agreement between the free field and Ward identity calculations only in the anomaly frame that was suited to the type of background field (gauge or metric) that was turned on. We argued that this was due to the presence of linearly growing potentials. This problem arises here as well, except that we now have no freedom to shift to an anomaly frame where the diffeomorphism anomaly is absent. We leave further investigation of this issue to later work. 

\section{Discussion}

This has been a long journey, and we now briefly review the path that we traveled. We studied the structure of entanglement entropy in quantum field theories with gravitational anomalies or mixed gauge-gravitational anomalies. Such theories are not precisely invariant under changes of coordinates, and we showed that indeed there is an {\it entanglement anomaly}, i.e. that the entanglement entropy in such theories changes in a well-defined manner under diffeomorphisms or U(1) gauge transformations.

We presented various derivations of this fact.  Beginning with two-dimensional conformal field theories, we reviewed an intuitive geometric arguments involving separation of left and right moving degrees of freedom as well as a (slightly) more formal discussion using conformal field theory techniques. In two and four dimensions we used a careful treatment of Ward identities to derive closed expressions for the diffeomorphism variation of the entanglement entropy as local integrals over the entangling surface. These results apply to any anomalous theory, conformal or not. In four dimensions the mixed anomaly can also be presented in a manner that preserves diffeomorphisms but breaks $U(1)$ charge conservation, and we showed that in this anomaly frame there is a closed expression for the $U(1)$ gauge variation of the entanglement entropy.

In four dimensions, for these expressions to be nonzero we must turn on background magnetic or gravitational fields. In four and six dimensions we then studied free fields (chiral fermions or self-dual bosons) moving in various background fields, permitting explicit calculations of the transformation of the entanglement entropy. The entanglement anomaly in sufficiently symmetric situations turns out to only be sensitive to the zero mode spectrum of chiral fermions; with the help of well-known index theorems we found precise agreement with the general formulas above. In the case of the 4d $U(1)$ gauge transformation this agreement involved a novel kind of ``zero'' mode for a Weyl fermion moving on a gravitational background with a nonzero ``twist'' turned on along the entangling surface.

We also studied another point of view on the entanglement anomaly by describing $d$-dimensional anomalous theories as living on the boundary of $d+1$-dimensional gapped ``Hall'' phases. These combined bulk + boundary systems are actually microscopically diffeomorphism-invariant, as the anomaly of the gapped bulk cancels that of the gapless boundary. Nevertheless, if we ``shear'' the boundary metric relative to the bulk, the entanglement entropy transforms in a universal manner that is completely captured by the anomaly. In principle the predictions of these sections could be verified by an direct computation on the wavefunction of the many-body state of electrons in a quantum Hall droplet, and there may be applications to condensed matter physics.

With the factual summary out of the way, we now turn to a discussion of the implications of our results. An immediate and seemingly obvious consequence is that even if a full theory is non-anomalous, if a subsector of it is anomalous then the entanglement of that subsector will be subject to the entanglement anomalies described above. For example, consider the Standard Model describing our universe.  If we sum over all the chiral fields then the mixed gauge-gravitational anomaly vanishes. However (in the presence of a magnetic field suitably coupling to $U(1)$ hypercharge), the entanglement entropy of {\it any one} of the fermion fields can only be defined up to the boost ambiguity described above! We find the idea that it is hard to define the entanglement of e.g. the electron neutrino field with the rest of the universe intriguing and feel it may lead to further insights.

There are also many more concrete directions for further study. The discussion of the relation between zero modes and entanglement anomalies is not yet complete. In particular, in six dimensions there are more cases to consider, including both the nonzero twist flux terms of the form \eqref{twistsquare}, as well as the physics of the mixed gauge-gravitational anomaly in six dimensions. It would also be interesting to further understand the field-theoretical implications of the novel ``zero'' mode on the twist field background. 

Finally, we find it interesting that the anomaly structure of a quantum field theory appears to be encoded into the entanglement structure of the many-body state. We hope that further study of the connections between entanglement and anomalies will better help us understand both of these fundamental ideas in quantum field theory.

\begin{acknowledgements}

It is a pleasure to acknowledge helpful discussions with H. Casini, A. Castro, D. Hofman, I. Ilgin, K. Jensen, Y. Liu, M. Metlitski, J. Kruthoff, W. Song, and E. Witten. We thank also T. Azeyanagai, G-S. Ng, T. Nishioka, R. Loganayagam and A. Yarom for correspondence and for sharing their drafts before publication. A.W. is grateful for support from the Simons Foundation, the Institute for Advanced Study, the KITP, and NSF grants No. PHY-1205500, PHY11-25915, and PHY-1314311 while this work was completed. N.I. is very grateful to the Perimeter Institute and the organizers of the KITPC Beijing workshop ``Holographic duality for condensed matter physics,'' for hospitality during the completion of this work. N.I. is supported by the D-ITP consortium, a program of the Netherlands Organisation for Scientific Research (NWO) that is funded by the Dutch Ministry of Education, Culture and Science (OCW).

\end{acknowledgements}

\begin{appendix}
\section{Geometric conventions} \label{app:geom}
Our conventions for analytic continuation are the same as in \cite{Castro:2014tta}, i.e. from Euclidean time $\tau_E$ to Lorentzian $t$ are connected by
\be
\tau_E = it \ .
\ee
We define the Christoffel symbol as
\be\label{eq:gg}
\Gamma^\alpha_{\mu\nu}={1\over 2} g^{\alpha\beta}\left[\partial_\mu g_{\nu\beta}+\partial_\nu g_{\mu\beta}-\partial_\beta g_{\mu\nu}\right]~,
\ee
and the Riemann tensor is
\be
R^\rho_{~\sigma \mu\nu}=\partial_\mu \Gamma^\rho_{\nu\sigma}-\partial_\nu \Gamma^\rho_{\mu\sigma} +\Gamma^\rho_{\mu\lambda}\Gamma^\lambda_{\nu\sigma}-\Gamma^\rho_{\nu\lambda}\Gamma^\lambda_{\mu\sigma} \ .
\ee

We will need some geometry of submanifolds. Some useful formulas are written below. A more detailed discussion can be found in \cite{poisson2004relativist,carter1992outer}. In the rest of this section we work entirely in Euclidean signature.

The entangling surface $\p A$ is a codimension-$2$ submanifold parametrized by coordinates $\sig^i$. If $x^{\mu}(\sig^i)$ are the embedding coordinates of the entangling surface, then the induced metric $h_{ij}$ on the entangling surface is
\be
h_{ij}(\sig) = g_{\mu\nu}(x(\sig))e^{\mu}_i e^{\nu}_j \qquad e^{\nu}_i \equiv \frac{\p X^{\nu}}{\p \sig^i} \label{indmet} \ .
\ee
 Consider the two normal vectors $n_a^{\mu}$ to this submanifold, where $a$ is an orthonormal index that runs over $1,2$, and we have $n^{a\mu} n^b_{\mu} = \delta^{ab}$. We construct projection tensors onto the entangling surface and its orthogonal complement respectively as:
 \be
 h_{\mu\nu} \equiv e^i_{\mu} e^{j}_{\nu} h_{ij} \qquad n_{\mu\nu} \equiv n^a_\mu n^b_\nu \delta_{ab},
 \ee
where the $i,j$ indices are raised and lowered using the induced metric from \eqref{indmet}. Now the usual extrinsic curvature is
 \be
 K_{aij} = K_{aji} \equiv \nabla_{\mu}n_{a\nu} e^{\mu}_i e^{\nu}_j
 \ee
The indices on the above object are a mix of tangential and normal; it is often helpful to think instead of a fully covariant extrinsic curvature tensor by defining
\be
K_{\mu\rho\sig} = K_{aij}n^{a}_{\mu} e^{i}_{\rho} e^{j}_{\sig} \label{covK}
\ee
We turn to geometric structures associated to the normal frame. Note that there is an $SO(2)$ gauge freedom associated with local rotations of this normal frame about the entangling surface:
\be
\delta n^{a}_{\mu} =  \Lam(\sig) \bar{\ep}^{ab}n^b_{\mu}
\ee
where $\bar{\ep}_{ab}$ is the $2$-dimensional antisymmetric symbol. We will call this degree of freedom the {\it twist}. We can define the binormal (a full spacetime tensor) to the entangling surface to be
\be
\ep^{\mu\nu}_{\p A} \equiv \bar{\ep}^{ab} n_{a}^{\mu}n_b^{\nu} \ . \label{binormal}
\ee
The binormal is invariant under the $SO(2)$ frame rotations discussed above. We can now define a twist $SO(2)$ gauge field that lives on the entangling surface:
\be
V^{ab}_{i} = -V_i^{ba} \equiv  n^a_{\rho} \nabla_{\mu} n^{b\rho} e^{\mu}_i
\ee
We can also define the field strength of of $U$ on the entangling surface as
\be
\Om_{ij}^{\phantom{ij}ab} \equiv \p_i V^{ab}_j - \p_j V^{ab}_i \label{curvature}
\ee
If the co-dimension of the sub-manifold in question is higher than $2$, then the twist symmetry is non-Abelian and there are extra terms present in \eqref{curvature} that we have omitted; we direct the reader to \cite{carter1992outer} for a complete treatment. We can define fully covariant versions of $V$ and $\Om$ as in \eqref{covK} via
\be
V^{\mu\nu}_{\rho} = V^{ab}_i n_{a}^{\mu} n_{b}^{\nu} e^i_{\rho}\qquad  \Om_{\mu\nu}^{\phantom{\mu\nu}\rho\sig} = \Om_{ij}^{\phantom{ij}ab}e^{i}_{\mu}e^{j}_{\nu}n_{a\rho} n_{b\sig} \ .
\ee
Now finally we present the Voss-Ricci relation between the twist field strength, the extrinsic curvature, and the background Riemann tensor:
\be
\Om_{\ka\lam}^{\phantom{\ka\lam}\mu\nu} = \le(K^{\mu}_{\phantom{\mu}\ka\sig} K^{\nu}_{\phantom{\nu}\lam\beta} - K^{\mu}_{\phantom{\mu}\lam\sig} K^{\nu}_{\phantom{\nu}\ka\beta}\ri) g^{\beta\sig} + h_{\ka}^{\rho}h_{\lam}^{\sig}n^{\mu}_{\tau}n^{\nu\al}R^{\tau}_{\phantom{\tau}\al\rho\sig} \label{vossricci}
\ee
This is an analog of the more familiar Gauss-Codazzi relations, which tell us how the extrinsic curvatures of a sub-manifold are related to its intrinsic curvature and to the background Riemann tensor with all or three indices projected tangentially to the sub-manifold. This less familiar relation projects the first two indices normally and the last two tangentially, and so only exists when there are at least two normal directions.

As $\Om_{\mu\nu\al\beta}$ and $V^{\mu\nu}_{\rho}$ are actually associated with a $U(1)$ gauge symmetry, there is no loss of information in contracting with the binormal to obtain simpler objects
\be
\Om_{\mu\nu} \equiv \ha \ep^{\al\beta}_{\p A}\Om_{\mu\nu\al\beta} \qquad V_{\mu} \equiv \ha \ep^{\p A}_{\al\beta} V^{\al\beta}_{\mu}  . \label{simpomega}
\ee

\section{Free fermions with twist flux} \label{app:fermions}
Here we present some details on the free fermion computations alluded to in the main text.
\subsection{Twist flux geometry and chiral modes}
We are interested in understanding the dynamics of Weyl fermions on the following background:
\be
ds^2 = dr^2 - r^2(d\eta - U_i(x^i) dx^i)^2 + dx^i dx^j \delta_{ij} + \cdots
\ee
where it is understood that the twist gauge field $U_i$ is small and we work only to first order in it. We work in $d = 2\mathbb{Z}$ dimensions, where we are interested in particular in $d = 4, 6$. $i,j$ run over an even dimensional transverse $\mathbb{R}^p$, $p = 2\mathbb{Z}$. The vielbein is defined as
\be
g_{\mu\nu} = e^{a}_{\mu} e^{b}_{\nu} \eta_{ab},
\ee
where in this section $a,b$ are tangent space indices and $\eta_{ab}$ is the flat Minkowski metric with mostly plus signature. Explicitly, the vielbeins are
\be
e^{\hat{r}} = dr \qquad e^{\hat{i}} = dx^i \qquad e^{\hat{\eta}} = r (d\eta - U_i dx^i) \ .
\ee
The spin connection can be computed from the torsion-free condition to be
\be
\om^{\hat{\eta}}_{\phantom{\eta}\hat{r}} = d\eta - \ha U_i dx^i \qquad \om^{\hat{\eta}}_{\phantom{\eta}\hat{i}}= - r \p_{[i}U_{j]} dx^j + \ha U_i dr \qquad \om^{\hat{i}}_{\phantom{i}\hat{j}} = -r^2 \p_{[i}U_{j]} d\eta \ .
\ee
We now seek to study the Weyl equation on this background. We will start with the Dirac equation in a chiral basis and eventually take a Weyl projection. We work with the following basis of Dirac matrices in $d$-dimensions:
\be
\qquad \Ga^{\eta} = \le(\begin{tabular}{c c}$0$ & $-1$ \\ $1$ & $0$\end{tabular}\ri) \qquad \Ga^{r} =  \le(\begin{tabular}{c c}$0$ & $\ga^{p+1}$ \\ $\ga^{p+1}$ & $0$\end{tabular}\ri) \qquad \Ga^{i} = \le(\begin{tabular}{c c}$0$ & $\ga^i$ \\ $\ga^i$ & $0$\end{tabular}\ri)
\ee
An explicit index on a gamma matrix should be understood as denoting a tangent space index, i.e. $\ga^{r} \equiv \ga^{\hat{r}}$, and thus there are no metric factors above. Here $\ga^i$ refers to a choice of Dirac matrices for $\mathbb{R}^p$ and $\ga^{p+1}$ refers to the chirality matrix for the $p$-dimensional spinor, i.e. if $p=4$ it is simply the usual $\ga^5$. If $p = 2$ we can take $\ga^i = \sig^i$ with $\sig^i$ the usual Pauli matrices, and $\ga^{p+1} = \sig^3$.

The Dirac action for a spinor in curved space is
\be
S = -\int d^4x \sqrt{-g} i \bpsi\;\ga^a e_{a}^{\mu} D_{\mu} \psi,
\ee
where the covariant derivative is
\be
D_{\mu} = \p_{\mu} + \frac{1}{8} \om_{ab,\mu} [\ga^{a},\ga^{b}] \ .
\ee
We may now directly compute the Dirac equation. In what follows we will work only with the equations of motion near the Rindler horizon $r \to 0$; thus we neglect all terms of the form $r\p_i U_j$, which are higher order in $r$, as well as all terms that are quadratic or higher in $U_i$. We find for the Dirac equation:
\be
\le[\le(\begin{tabular}{c c}$0$ & $\ga^{p+1}$ \\ $\ga^{p+1}$ & $0$\end{tabular}\ri)\le(\p_r + \frac{1}{2r}\ri) + \frac{1}{r}\le(\begin{tabular}{c c}$0$ & $-1$ \\ $1$ & $0$\end{tabular}\ri)\p_{\eta} + \le(\begin{tabular}{c c}$0$ & $D_{\perp}$ \\ $D_{\perp}$ & $0$\end{tabular}\ri)\ri] \psi = 0.
\ee
where the operator $D_{\perp}$ on the transverse space is:
\be
D_{\perp} \equiv \ga^i\le(\p_i + U_i \p_{\eta}\ri)
\ee
We see that this looks like the coupling of a Dirac spinor to a $U(1)$ gauge field $U_i$ in the transverse space. This can be made more precise if we work in frequency space, giving all fields time dependence of the form $e^{-i \om \eta}$; then the effective $U(1)$ charge of each of these modes is simply the Rindler energy $\om$.

We now finally take a Weyl projection to find that the Weyl equation in $d$-dimensions is
\be
\le(\pm \p_{\eta} + \ga^{p+1}\le(r \p_r + \ha \ri) + r D_{\perp}\ri)\psi_{R,L} = 0
\ee
We see that the spectrum of $D_{\perp}$ is crucial in determining the low-energy properties of the Rindler modes. In particular, a zero mode of $D_{\perp}$ with a definite chirality under $\ga^{p+1}$ will map to a mode in Rindler space with a chiral nature.

We can now make contact with the usual index theorems for $U(1)$ gauge fields coupled to Dirac fermions in $p$-dimensions. We should note that this is actually somewhat heuristic. Normally index theorems apply to compact manifolds; however we cannot really compactify the $\mathbb{R}^p$ in a sensible way, since the $U(1)$ gauge symmetry in question is not compact\footnote{In other words, Rindler energy is not quantized: it is more appropriate to call $U_i$ a $\mathbb{R}$ gauge field.}. Nevertheless we will obtain sensible answers.

We start in $p=2$. Then the 2d index theorem tells us that the numbers of definite chirality zero modes of $D_{\perp}$ satisfy:
\be
N_+ - N_- = \frac{\om}{2\pi} \int d^2x\;dU \label{Rindlerindex}
\ee
As none of the quantities on the right-hand side are quantized, this should be thought of as characterizing the {\it density} of zero modes per unit-$p$ volume. In $p=2$ we see that the net chirality of the modes is determined by the sign of the Rindler energy. This is somewhat novel, resulting in a chiral charge flow in the Rindler directions and a resulting gauge anomaly in the entanglement entropy. We discuss its implications in detail in the main text and below.

We now move to $p = 4$, though we do not actually use it in this paper. Now the 4d index theorem tells us that
\be
N_+ - N_- = \frac{\om^2}{32\pi^2} \int d^4x \ep^{ijkl}(dU)_{ij}(dU)_{kl}
\ee
We see that in this case it is $\om^2$ that appears on the right-hand side and not $\om$. Thus the net chirality of the zero modes is determined by the sign of $dU \wedge dU$ and not by the sign of the Rindler energy. This is more conventional then the case above. Each of these zero modes depends on the Rindler coordinates as an ordinary 2d Weyl fermion, presumably resulting in the diffeomorphism anomaly in the entanglement entropy discussed in \eqref{twistsquare}. 

The fact that the {\it degeneracy} of the modes depends on the Rindler energy does indicate that the physics in the Rindler directions is still somewhat exotic.

\subsection{Canonical quantization of free fermion modes}
In this section we provide a discussion of the physics arising from \eqref{Rindlerindex}. As discussed in the main text, if we build a full spinor from $\psi_L = \chi^{\pm}(x,y)\Psi_{\om}^{\pm}(r)e^{-i\om\eta}$, we find for the equation of motion in the Rindler radial coordinate:
\begin{align}
\le(r\p_r + \ha + i \om\ri)\Psi_{\om}^{+}(r) & = 0,\qquad \om\sB > 0 \\
\le(r\p_r + \ha - i \om\ri)\Psi_{\om}^{-}(r) & = 0,\qquad \om\sB < 0
\end{align}
For concreteness, let us now fix the sign of $\sB > 0$. We then find that
\be
\Psi^{\pm}_{\om}(r) \sim \frac{1}{\sqrt{r}} e^{-i|\om| \log r} \label{psiexp}
\ee
In other words, independent of the sign of $\om$, the spatial momentum in the $r$-direction always has the same sign. This is somewhat novel, and we devote the rest of this appendix to the canonical quantization of such a $(1+1)d$ system. The ultimate goal is to derive the expression for the current \eqref{lorj} in the main text.

We work only with the interesting zero mode sector, suppressing the transverse directions (which provide only a constant density of states). Due to the restriction on spatial momenta present in \eqref{psiexp}, the fermion field may be expanded in terms of only positive momentum modes as
\be
\psi(r) = \int_{p>0} \frac{dp}{2\pi}\frac{e^{-i p \log r}}{\sqrt{r}}\le(a_p v_2 + b_p^{\dagger} v_1\ri)
\ee
where $v_1 = (1,0)^T$, $v_2 = (0,1)^T$. $a_p$ and $b_p$ are the raising and lowering operators for those modes. The Weyl action turns out to be:
\be
S = i\int dr d\eta \le( \psi^{\dagger} \p_{\eta} \psi + r \psi^{\dagger}\sig^3 \p_r \psi\ri) \ .
\ee
Thus the canonical momentum to $\psi$ is
\be
\pi_{\psi} = i \psi^{\dagger},
\ee
leading to the canonical quantization condition
\be
\{\psi^{\dagger}(r), \psi(r')\} = \delta(r - r')
\ee
Let us now use this to determine the canonical commutation relations of $a_p$ and $b_p$. We may extract $a_p$ from $\psi$ as follows:
\be
a_p = v_2^{\dagger}\int \frac{dr}{\sqrt{r}} e^{i p \log r} \psi(r),
\ee
and similarly for $b_q$. We then find that
\be
\{a_p, a^{\dagger}_{p'}\} = (2\pi) \delta(p-p') \qquad \{b_p, b^{\dagger}_{p'}\} = (2\pi) \delta(p-p')
\ee
Thus these behave as normal fermion raising and lowering operators. Note that so far the interesting restriction to positive momenta has not played a role.

Let us now find the Hamiltonian so that we determine the correct vacuum. The Hamiltonian is
\be
H = \pi \p_{\eta}\psi - \sL =  -i \int dr r \psi^{\dagger}\sig^3 \p_r \psi, \label{ham}
\ee
which can be worked out in modes to be
\be
H = \int \frac{dr}{r} \frac{dp dp'}{(2\pi)^2}\le(a_p^{\dagger} a_p' - b_p b_p'^{\dagger}\ri)p' e^{(i \log r)(q - p)}
\ee
The relative minus sign between the two sets of oscillators arises from the $\sig^3$ in the Hamiltonian in \eqref{ham}. We now anti-commute the $b$'s, neglect the zero-point energy, and do some integrals to find
\be
H = \int_{p>0} \frac{dp}{2\pi} p\le(a_p^{\dagger}a_p + b_p^{\dagger} b_p\ri)
\ee
This shows that we may define the vacuum by $a_p|0\rangle = 0$, $b_p|0\rangle = 0$. The Hamiltonian is positive-definite because of the restriction to positive momenta.

We now turn to the current. The original charge current is defined as
\be
j^{\mu} = -q\bpsi \ga^{\mu} \psi
\ee
Thus the contribution to the radial current from the zero-point sector is (up to zero-point contributions, which we neglect as they do not contribute to the entanglement anomaly in question):
\be
j^r(r) = q\int \frac{dp dp'}{(2\pi)^2} \le(a_p^{\dagger}a_{p'} + b_{p'}^{\dagger}b_p\ri)\frac{e^{i \log r(q-p)}}{r} \ .
\ee
Note both particles and anti-particles contribute with the same sign to the current, as one might have expected from heuristic considerations described in the main text. Now in the thermal state the density matrix is diagonal in a momentum basis, and we have
\be
\langle b_{p'}^{\dagger}b_p\rangle = \langle a_p^{\dagger} a_{p'} \rangle = (2\pi) \delta(p - p') n(\om_p) \qquad n(\om) \equiv \frac{1}{1 + e^{\beta\om}}
\ee
where for us $\om_p = p$ and $n(\om)$ is simply the Fermi-Dirac distribution. This gives us the expression for the current used in \eqref{lorj}.

\end{appendix}

\bibliography{entano9}{}

\providecommand{\href}[2]{#2}\begingroup\raggedright\begin{thebibliography}{10}

\bibitem{Calabrese:2005zw}
P.~Calabrese and J.~L. Cardy, {\it {Entanglement entropy and quantum field
  theory: A Non-technical introduction}},  {\em Int.J.Quant.Inf.} {\bf 4}
  (2006) 429, [\href{http://xxx.lanl.gov/abs/quant-ph/0505193}{{\tt
  quant-ph/0505193}}].

\bibitem{Amico:2007ag}
L.~Amico, R.~Fazio, A.~Osterloh, and V.~Vedral, {\it {Entanglement in many-body
  systems}},  {\em Rev. Mod. Phys.} {\bf 80} (2008) 517--576,
  [\href{http://xxx.lanl.gov/abs/quant-ph/0703044}{{\tt quant-ph/0703044}}].

\bibitem{Ryu:2006bv}
S.~Ryu and T.~Takayanagi, {\it {Holographic derivation of entanglement entropy
  from AdS/CFT}},  {\em Phys. Rev. Lett.} {\bf 96} (2006) 181602,
  [\href{http://xxx.lanl.gov/abs/hep-th/0603001}{{\tt hep-th/0603001}}].

\bibitem{Solodukhin:2011gn}
S.~N. Solodukhin, {\it {Entanglement entropy of black holes}},  {\em Living
  Rev. Rel.} {\bf 14} (2011) 8, [\href{http://xxx.lanl.gov/abs/1104.3712}{{\tt
  arXiv:1104.3712}}].

\bibitem{Wall:2011kb}
A.~C. Wall, {\it {Testing the Generalized Second Law in 1+1 dimensional
  Conformal Vacua: An Argument for the Causal Horizon}},  {\em Phys.Rev.} {\bf
  D85} (2012) 024015, [\href{http://xxx.lanl.gov/abs/1105.3520}{{\tt
  arXiv:1105.3520}}].

\bibitem{Castro:2014tta}
A.~Castro, S.~Detournay, N.~Iqbal, and E.~Perlmutter, {\it {Holographic
  entanglement entropy and gravitational anomalies}},  {\em JHEP} {\bf 07}
  (2014) 114, [\href{http://xxx.lanl.gov/abs/1405.2792}{{\tt
  arXiv:1405.2792}}].

\bibitem{Loga}
T.~Azeyanagi, R.~Loganayagam, and G.~S. Ng, {\it {Holographic Entanglement for
  Chern-Simons Terms}},  \href{http://xxx.lanl.gov/abs/1507.0229}{{\tt
  arXiv:1507.0229}}.

\bibitem{NY}
T.~Nishioka and A.~Yarom, {\it {Anomalies and Entanglement Entropy}},  {\em to
  appear}.

\bibitem{Guo:2015uqa}
W.-z. Guo and R.-x. Miao, {\it {Entropy for gravitational Chern-Simons terms by
  squashed cone method}},  \href{http://xxx.lanl.gov/abs/1506.0839}{{\tt
  arXiv:1506.0839}}.

\bibitem{Hosseini:2015uba}
S.~M. Hosseini and A.~Veliz-Osorio, {\it {Gravitational anomalies, entanglement
  entropy, and flat-space holography}},
  \href{http://xxx.lanl.gov/abs/1507.0662}{{\tt arXiv:1507.0662}}.

\bibitem{Belin:2015jpa}
A.~Belin, A.~Castro, and L.-Y. Hung, {\it {Fake gaps in AdS3/CFT2}},
  \href{http://xxx.lanl.gov/abs/1508.0120}{{\tt arXiv:1508.0120}}.

\bibitem{Fursaev:2013fta}
D.~V. Fursaev, A.~Patrushev, and S.~N. Solodukhin, {\it {Distributional
  Geometry of Squashed Cones}},  {\em Phys. Rev.} {\bf D88} (2013), no.~4
  044054, [\href{http://xxx.lanl.gov/abs/1306.4000}{{\tt arXiv:1306.4000}}].

\bibitem{Solodukhin:2008dh}
S.~N. Solodukhin, {\it {Entanglement entropy, conformal invariance and
  extrinsic geometry}},  {\em Phys. Lett.} {\bf B665} (2008) 305--309,
  [\href{http://xxx.lanl.gov/abs/0802.3117}{{\tt arXiv:0802.3117}}].

\bibitem{Holzhey:1994we}
C.~Holzhey, F.~Larsen, and F.~Wilczek, {\it {Geometric and renormalized entropy
  in conformal field theory}},  {\em Nucl. Phys.} {\bf B424} (1994) 443--467,
  [\href{http://xxx.lanl.gov/abs/hep-th/9403108}{{\tt hep-th/9403108}}].

\bibitem{Ryu:2006ef}
S.~Ryu and T.~Takayanagi, {\it {Aspects of Holographic Entanglement Entropy}},
  {\em JHEP} {\bf 08} (2006) 045,
  [\href{http://xxx.lanl.gov/abs/hep-th/0605073}{{\tt hep-th/0605073}}].

\bibitem{Hung:2011xb}
L.-Y. Hung, R.~C. Myers, and M.~Smolkin, {\it {On Holographic Entanglement
  Entropy and Higher Curvature Gravity}},  {\em JHEP} {\bf 04} (2011) 025,
  [\href{http://xxx.lanl.gov/abs/1101.5813}{{\tt arXiv:1101.5813}}].

\bibitem{Schwimmer:2008yh}
A.~Schwimmer and S.~Theisen, {\it {Entanglement Entropy, Trace Anomalies and
  Holography}},  {\em Nucl. Phys.} {\bf B801} (2008) 1--24,
  [\href{http://xxx.lanl.gov/abs/0802.1017}{{\tt arXiv:0802.1017}}].

\bibitem{Casini:2011kv}
H.~Casini, M.~Huerta, and R.~C. Myers, {\it {Towards a derivation of
  holographic entanglement entropy}},  {\em JHEP} {\bf 05} (2011) 036,
  [\href{http://xxx.lanl.gov/abs/1102.0440}{{\tt arXiv:1102.0440}}].

\bibitem{AlvarezGaume:1983ig}
L.~Alvarez-Gaume and E.~Witten, {\it {Gravitational Anomalies}},  {\em Nucl.
  Phys.} {\bf B234} (1984) 269.

\bibitem{Buividovich:2008gq}
P.~V. Buividovich and M.~I. Polikarpov, {\it {Entanglement entropy in gauge
  theories and the holographic principle for electric strings}},  {\em Phys.
  Lett.} {\bf B670} (2008) 141--145,
  [\href{http://xxx.lanl.gov/abs/0806.3376}{{\tt arXiv:0806.3376}}].

\bibitem{Donnelly:2011hn}
W.~Donnelly, {\it {Decomposition of entanglement entropy in lattice gauge
  theory}},  {\em Phys. Rev.} {\bf D85} (2012) 085004,
  [\href{http://xxx.lanl.gov/abs/1109.0036}{{\tt arXiv:1109.0036}}].

\bibitem{Casini:2013rba}
H.~Casini, M.~Huerta, and J.~A. Rosabal, {\it {Remarks on entanglement entropy
  for gauge fields}},  {\em Phys. Rev.} {\bf D89} (2014), no.~8 085012,
  [\href{http://xxx.lanl.gov/abs/1312.1183}{{\tt arXiv:1312.1183}}].

\bibitem{'tHooft:1984re}
G.~'t~Hooft, {\it {On the Quantum Structure of a Black Hole}},  {\em Nucl.
  Phys.} {\bf B256} (1985) 727.

\bibitem{Casini:2004bw}
H.~Casini and M.~Huerta, {\it {A Finite entanglement entropy and the
  c-theorem}},  {\em Phys. Lett.} {\bf B600} (2004) 142--150,
  [\href{http://xxx.lanl.gov/abs/hep-th/0405111}{{\tt hep-th/0405111}}].

\bibitem{Calabrese:2004eu}
P.~Calabrese and J.~L. Cardy, {\it {Entanglement entropy and quantum field
  theory}},  {\em J.Stat.Mech.} {\bf 0406} (2004) P06002,
  [\href{http://xxx.lanl.gov/abs/hep-th/0405152}{{\tt hep-th/0405152}}].

\bibitem{Cardy:2007mb}
J.~Cardy, O.~Castro-Alvaredo, and B.~Doyon, {\it {Form factors of branch-point
  twist fields in quantum integrable models and entanglement entropy}},  {\em
  J.Statist.Phys.} {\bf 130} (2008) 129--168,
  [\href{http://xxx.lanl.gov/abs/0706.3384}{{\tt arXiv:0706.3384}}].

\bibitem{Nelson:1994na}
W.~Nelson, {\it {A Comment on black hole entropy in string theory}},  {\em
  Phys. Rev.} {\bf D50} (1994) 7400--7402,
  [\href{http://xxx.lanl.gov/abs/hep-th/9406011}{{\tt hep-th/9406011}}].

\bibitem{Ohmori:2014eia}
K.~Ohmori and Y.~Tachikawa, {\it {Physics at the entangling surface}},  {\em J.
  Stat. Mech.} {\bf 1504} (2015), no.~4 P04010,
  [\href{http://xxx.lanl.gov/abs/1406.4167}{{\tt arXiv:1406.4167}}].

\bibitem{Donnelly:2014fua}
W.~Donnelly and A.~C. Wall, {\it {Entanglement entropy of electromagnetic edge
  modes}},  {\em Phys. Rev. Lett.} {\bf 114} (2015), no.~11 111603,
  [\href{http://xxx.lanl.gov/abs/1412.1895}{{\tt arXiv:1412.1895}}].

\bibitem{Donnelly:2015hxa}
W.~Donnelly and A.~C. Wall, {\it {Geometric entropy and edge modes of the
  electromagnetic field}},  \href{http://xxx.lanl.gov/abs/1506.0579}{{\tt
  arXiv:1506.0579}}.

\bibitem{Donnelly:2014gva}
W.~Donnelly, {\it {Entanglement entropy and nonabelian gauge symmetry}},  {\em
  Class. Quant. Grav.} {\bf 31} (2014), no.~21 214003,
  [\href{http://xxx.lanl.gov/abs/1406.7304}{{\tt arXiv:1406.7304}}].

\bibitem{Huang:2014pfa}
K.-W. Huang, {\it {Central Charge and Entangled Gauge Fields}},  {\em Phys.
  Rev.} {\bf D92} (2015), no.~2 025010,
  [\href{http://xxx.lanl.gov/abs/1412.2730}{{\tt arXiv:1412.2730}}].

\bibitem{Nielsen:1980rz}
H.~B. Nielsen and M.~Ninomiya, {\it {Absence of Neutrinos on a Lattice. 1.
  Proof by Homotopy Theory}},  {\em Nucl. Phys.} {\bf B185} (1981) 20.
  [Erratum: Nucl. Phys.B195,541(1982)].

\bibitem{Nielsen:1981xu}
H.~B. Nielsen and M.~Ninomiya, {\it {Absence of Neutrinos on a Lattice. 2.
  Intuitive Topological Proof}},  {\em Nucl. Phys.} {\bf B193} (1981) 173.

\bibitem{wen1992theory}
X.-G. Wen, {\it Theory of the edge states in fractional quantum hall effects},
  {\em International Journal of Modern Physics B} {\bf 6} (1992), no.~10
  1711--1762.

\bibitem{fradkin2013field}
E.~Fradkin, {\em Field theories of condensed matter physics}.
\newblock Cambridge University Press, 2013.

\bibitem{Schellekens:1992db}
A.~N. Schellekens, {\it {Meromorphic C = 24 conformal field theories}},  {\em
  Commun. Math. Phys.} {\bf 153} (1993) 159--186,
  [\href{http://xxx.lanl.gov/abs/hep-th/9205072}{{\tt hep-th/9205072}}].

\bibitem{Calabrese:2009qy}
P.~Calabrese and J.~Cardy, {\it {Entanglement entropy and conformal field
  theory}},  {\em J.Phys.} {\bf A42} (2009) 504005,
  [\href{http://xxx.lanl.gov/abs/0905.4013}{{\tt arXiv:0905.4013}}].

\bibitem{Bardeen:1984pm}
W.~A. Bardeen and B.~Zumino, {\it {Consistent and Covariant Anomalies in Gauge
  and Gravitational Theories}},  {\em Nucl.Phys.} {\bf B244} (1984) 421.

\bibitem{AlvarezGaume:1984dr}
L.~Alvarez-Gaume and P.~H. Ginsparg, {\it {The Structure of Gauge and
  Gravitational Anomalies}},  {\em Annals Phys.} {\bf 161} (1985) 423.
  [Erratum: Annals Phys.171,233(1986)].

\bibitem{Ginsparg:1985qn}
P.~H. Ginsparg, {\it {Applications of Topological and Differential Geometric
  Methods to Anomalies in Quantum Field Theory}}, .

\bibitem{Jensen:2012kj}
K.~Jensen, R.~Loganayagam, and A.~Yarom, {\it {Thermodynamics, gravitational
  anomalies and cones}},  {\em JHEP} {\bf 02} (2013) 088,
  [\href{http://xxx.lanl.gov/abs/1207.5824}{{\tt arXiv:1207.5824}}].

\bibitem{Jensen:2013kka}
K.~Jensen, R.~Loganayagam, and A.~Yarom, {\it {Anomaly inflow and thermal
  equilibrium}},  {\em JHEP} {\bf 05} (2014) 134,
  [\href{http://xxx.lanl.gov/abs/1310.7024}{{\tt arXiv:1310.7024}}].

\bibitem{Jensen:2013rga}
K.~Jensen, R.~Loganayagam, and A.~Yarom, {\it {Chern-Simons terms from thermal
  circles and anomalies}},  {\em JHEP} {\bf 05} (2014) 110,
  [\href{http://xxx.lanl.gov/abs/1311.2935}{{\tt arXiv:1311.2935}}].

\bibitem{Kraus:2005zm}
P.~Kraus and F.~Larsen, {\it {Holographic gravitational anomalies}},  {\em
  JHEP} {\bf 01} (2006) 022,
  [\href{http://xxx.lanl.gov/abs/hep-th/0508218}{{\tt hep-th/0508218}}].

\bibitem{2012PhRvB..85r4503S}
M.~{Stone}, {\it {Gravitational anomalies and thermal Hall effect in
  topological insulators}},  {\em \prb} {\bf 85} (May, 2012) 184503,
  [\href{http://xxx.lanl.gov/abs/1201.4095}{{\tt arXiv:1201.4095}}].

\bibitem{Deser:1981wh}
S.~Deser, R.~Jackiw, and S.~Templeton, {\it {Topologically Massive Gauge
  Theories}},  {\em Annals Phys.} {\bf 140} (1982) 372--411. [Annals
  Phys.281,409(2000)].

\bibitem{Fursaev:1995ef}
D.~V. Fursaev and S.~N. Solodukhin, {\it {On the description of the Riemannian
  geometry in the presence of conical defects}},  {\em Phys. Rev.} {\bf D52}
  (1995) 2133--2143, [\href{http://xxx.lanl.gov/abs/hep-th/9501127}{{\tt
  hep-th/9501127}}].

\bibitem{Fukushima:2008xe}
K.~Fukushima, D.~E. Kharzeev, and H.~J. Warringa, {\it {The Chiral Magnetic
  Effect}},  {\em Phys. Rev.} {\bf D78} (2008) 074033,
  [\href{http://xxx.lanl.gov/abs/0808.3382}{{\tt arXiv:0808.3382}}].

\bibitem{Swingle:2010bt}
B.~Swingle, {\it {Highly entangled quantum systems in 3+1 dimensions}},
  \href{http://xxx.lanl.gov/abs/1003.2434}{{\tt arXiv:1003.2434}}.

\bibitem{Landsteiner:2011cp}
K.~Landsteiner, E.~Megias, and F.~Pena-Benitez, {\it {Gravitational Anomaly and
  Transport}},  {\em Phys. Rev. Lett.} {\bf 107} (2011) 021601,
  [\href{http://xxx.lanl.gov/abs/1103.5006}{{\tt arXiv:1103.5006}}].

\bibitem{Son:2009tf}
D.~T. Son and P.~Surowka, {\it {Hydrodynamics with Triangle Anomalies}},  {\em
  Phys. Rev. Lett.} {\bf 103} (2009) 191601,
  [\href{http://xxx.lanl.gov/abs/0906.5044}{{\tt arXiv:0906.5044}}].

\bibitem{Dong:2013qoa}
X.~Dong, {\it {Holographic Entanglement Entropy for General Higher Derivative
  Gravity}},  {\em JHEP} {\bf 01} (2014) 044,
  [\href{http://xxx.lanl.gov/abs/1310.5713}{{\tt arXiv:1310.5713}}].

\bibitem{Camps:2013zua}
J.~Camps, {\it {Generalized entropy and higher derivative Gravity}},  {\em
  JHEP} {\bf 03} (2014) 070, [\href{http://xxx.lanl.gov/abs/1310.6659}{{\tt
  arXiv:1310.6659}}].

\bibitem{Landsteiner:2011iq}
K.~Landsteiner, E.~Megias, L.~Melgar, and F.~Pena-Benitez, {\it {Holographic
  Gravitational Anomaly and Chiral Vortical Effect}},  {\em JHEP} {\bf 09}
  (2011) 121, [\href{http://xxx.lanl.gov/abs/1107.0368}{{\tt
  arXiv:1107.0368}}].

\bibitem{ChristensenDuff79}
S.~M. Christensen and M.~J. Duff, {\it {New Gravitational Index Theorems and
  Supertheorems}},  {\em Nucl. Phys.} {\bf B154} (1979) 301.

\bibitem{poisson2004relativist}
E.~Poisson, {\em A Relativist's Toolkit: The Mathematics of Black-Hole
  Mechanics}.
\newblock Cambridge University Press, 2004.

\bibitem{carter1992outer}
B.~Carter, {\it Outer curvature and conformal geometry of an imbedding},  {\em
  Journal of Geometry and Physics} {\bf 8} (1992), no.~1 53--88.

\end{thebibliography}\endgroup
\bibliographystyle{JHEP}

\end{document}